\documentclass[sigconf,10pt]{acmart}
\renewcommand\footnotetextcopyrightpermission[1]{} 
\makeatletter
\renewcommand\@formatdoi[1]{\ignorespaces}
\makeatother

\AtBeginDocument{%
  \providecommand\BibTeX{{%
    \normalfont B\kern-0.5em{\scshape i\kern-0.25em b}\kern-0.8em\TeX}}}

\setcopyright{acmcopyright}
\copyrightyear{}
\acmYear{}
\acmDOI{}

\acmConference[]{}{}{}

\acmPrice{}
\acmISBN{}

\usepackage{xspace}
\usepackage{algorithmicx}
\usepackage{subcaption,graphicx,url,multirow}
\usepackage{algorithm}
\usepackage[noend]{algpseudocode}

\algrenewcommand\algorithmicindent{0.8em}

\algnewcommand\algorithmicparfor{\textbf{par-for}}
\algdef{S}[FOR]{ParFor}[1]{\algorithmicparfor\ #1\ \algorithmicdo}
\algnewcommand\algorithmicparforeach{\textbf{par-for each}}
\algdef{S}[FOR]{ParForEach}[1]{\algorithmicparforeach\ #1\ \algorithmicdo}
\algnewcommand\algorithmicforeach{\textbf{for each}}
\algdef{S}[FOR]{ForEach}[1]{\algorithmicforeach\ #1\ \algorithmicdo}

\usepackage{balance}
\usepackage{paralist}
\usepackage{booktabs}
\usepackage{color}
\usepackage{arydshln}
\usepackage{enumitem}
\usepackage{xcolor,soul}

\usepackage{soul,color}
\soulregister\cite7
\soulregister\ref7
\soulregister\pageref7
\soulregister\textproc7
\soulregister\textit7
\soulregister\defn7
\soulregister\,7

\newcommand{\highlight}[1]{{#1}}

\newcommand{\myparagraph}[1]{\medskip\noindent {\bf #1.}}

\newcommand{\codevar}[1]{\mathit{#1}}
\newcommand{\minpts}{\textsf{minPts}}

\newcommand{\hide}[1]{} 
\newcommand{\mb}[1]{{\mbox{\emph{#1}}}}

\usepackage{amsmath,amsthm,mathrsfs}

\newcommand{\whp}[1]{\emph{whp}}
\newcommand{\defn}[1]{{{\bf{\emph{#1}}}}}

\newcommand{\clusters}{\mb{clusters}}

\newcommand{\coreFlags}{\mb{coreFlags}}
\newcommand{\arrcount}{\mb{count}}

\let \originalleft \left
\let\originalright\right
\renewcommand{\left}{\mathopen{}\mathclose\bgroup\originalleft}
\renewcommand{\right}{\aftergroup\egroup\originalright}

\hide{
\setlength{\textfloatsep}{2pt plus 1.0pt minus 2.0pt}
\setlength{\intextsep}{2pt plus 1.0pt minus 2.0pt}
\setlength{\floatsep}{2pt plus 1.0pt minus 2.0pt}
\setlength{\dbltextfloatsep}{2pt plus 1.0pt minus 2.0pt}
\setlength{\dblfloatsep}{2pt plus 1.0pt minus 2.0pt}

\usepackage[compact]{titlesec}
\titlespacing*{\section}{0pt}{2pt}{2pt}
\titlespacing*{\subsection}{0pt}{1pt}{1pt}
\setlength{\textfloatsep}{1pt} 
}

\begin{document}
\fancyhead{}

\title[Theoretically-Efficient and Practical Parallel DBSCAN]{Theoretically-Efficient and Practical\\Parallel DBSCAN}
\titlenote{This is the full version of the paper appearing in the ACM SIGMOD International Conference on Management of Data (SIGMOD), 2020.}




\author{Yiqiu Wang}
\email{yiqiuw@mit.edu}
\affiliation{%
  \institution{MIT CSAIL}
}

\author{Yan Gu}
\email{ygu@ucr.edu}
\affiliation{%
  \institution{UC Riverside}
}

\author{Julian Shun}
\email{jshun@mit.edu}
\affiliation{%
  \institution{MIT CSAIL}
}

\begin{abstract}
  The DBSCAN method for spatial clustering has received significant
  attention due to its applicability in a variety of data analysis
  tasks.  There are fast sequential algorithms for DBSCAN in Euclidean
  space that take $O(n\log n)$ work for two dimensions, sub-quadratic
  work for three or more dimensions, and can be computed approximately
  in linear work for any constant number of dimensions.  However,
  existing parallel DBSCAN algorithms require quadratic work in the
  worst case.  This paper
  bridges the gap between theory and practice of parallel DBSCAN by
  presenting new parallel algorithms for Euclidean exact DBSCAN and
  approximate DBSCAN that match the work bounds of their sequential
  counterparts, and are highly parallel (polylogarithmic depth). We
  present implementations of our algorithms along with optimizations
  that improve their practical performance.  We perform a
  comprehensive experimental evaluation of our algorithms on a variety
  of datasets and parameter settings. Our experiments on a 36-core
  machine with two-way hyper-threading show that our implementations
  outperform existing parallel implementations by up to several orders
  of magnitude, and achieve speedups of up to 33x over the best
  sequential algorithms.
\end{abstract}

\maketitle

\section{Introduction}\label{sec:intro}

Spatial clustering methods are frequently used  to
group together similar objects. Density-based spatial clustering of
applications with noise (DBSCAN) is a popular method developed by Ester et
al.~\cite{Ester1996} that is able to find good clusters of
different shapes in the presence of noise without requiring prior
knowledge of the number of clusters. The DBSCAN algorithm has been
applied successfully to clustering in spatial databases, with
applications in various domains such as transportation, biology, and
astronomy.

The traditional DBSCAN algorithm~\cite{Ester1996} and their variants
require work quadratic in the input size in the worst case, which can
be prohibitive for the large data sets that need to be analyzed today.
To address this computational bottleneck, there has been recent work
on designing parallel algorithms for DBSCAN and its
variants~\cite{BrecheisenKP06,Xu1999,Arlia2001,Januzaj2004,Januzaj2004a,Bohm2009,Fu2011,Huang2017,Coppola2002,Patwary12,Patwary2013,Patwary2014,Patwary2015,Welton2013,Lulli2016,Fu2014,Dai2012,Andrade2013,Han2016,Kim2014,He2014,Fu2014,Chen2015,Gotz2015,Cordova2015,Luo2016,Yu2015,Hu2017,Araujo2015,Hu2018,Song2018}. However,
even though these solutions achieve scalability and speedup over
sequential algorithms, in the worst-case their number of operations
still scale quadratically with the input size.
Therefore, a natural
question is whether there exist DBSCAN algorithms that are faster
both in theory and practice, and in both the sequential and parallel
settings.

Given the ubiquity of datasets in Euclidean space, there has been
work on faster sequential DBSCAN algorithms in this setting.
Gunawan~\cite{Gunawan13} and de Berg et al.~\cite{BergGR17} has shown
that Euclidean DBSCAN in 2D can be solved sequentially in
$O(n\log n)$ work. Gan and Tao~\cite{GanT17} provide alternative
Euclidean DBSCAN algorithms for two-dimensions that take $O(n\log n)$
work. For higher dimensions, Chen et al.~\cite{Chen05} provide an
algorithm that takes $O(n^{2(1-1/(d+2))}\mbox{polylog}(n))$ work for
$d$ dimensions, and Gan and Tao~\cite{GanT17} improve the result with
an algorithm that takes $O(n^{2-(2/(\lceil d/2 \rceil +1))+\delta})$
work for any constant $\delta>0$.  To further reduce the work complexity, there
have been approximate DBSCAN algorithms proposed. Chen et
al.~\cite{Chen05} provide an approximate DBSCAN algorithm that takes
$O(n\log n)$ work for any constant number of dimensions, and Gan and
Tao~\cite{GanT17} provide a similar algorithm taking $O(n)$ expected
work.  However, none of the algorithms described above have been
parallelized.

This paper bridges the gap between theory and practice in
parallel Euclidean DBSCAN  by providing new parallel
algorithms for exact and approximate DBSCAN with work complexity
matching that of best sequential
algorithms~\cite{Gunawan13,BergGR17,GanT17}, and with low depth, which is the gold standard in parallel
algorithm design.
For exact 2D DBSCAN, we
design several parallel algorithms that use either the box or the grid
construction for partitioning points~\cite{Gunawan13,BergGR17} and one of the following three procedures
for determining connectivity among core points: Delaunay
triangulation~\cite{GanT17}, unit-spherical emptiness
checking with line separation~\cite{GanT17}, and bichromatic closest pairs.  For
higher-dimensional exact DBSCAN, we provide an algorithm based
on solving the higher-dimensional bichromatic closest pairs problem in
parallel. Unlike many existing parallel algorithms, our exact
algorithms produce the same results according to the standard definition of
DBSCAN, and so we do not sacrifice clustering quality.  For
approximate DBSCAN, we design an algorithm that uses parallel
quadtree construction and querying.  Our approximate algorithm
returns the same result as the sequential approximate algorithm by Gan and
Tao~\cite{GanT17}.



We perform a comprehensive set of experiments on  synthetic and
real-world datasets using varying parameters, and compare our
performance to optimized sequential implementations as well as
existing parallel DBSCAN algorithms.
On a 36-core machine with
two-way hyper-threading, our exact DBSCAN implementations achieve 2--89x
(24x on average) self-relative speedup  and 5--33x (16x on average) speedup over the fastest sequential
implementations. Our approximate DBSCAN implementations achieve 14--44x (24x on average)
self-relative speedup.
Compared to existing parallel
algorithms, which are scalable but have high overheads compared to serial implementations, our fastest exact algorithms are faster by up to orders of magnitude (16--6102x) under correctly chosen parameters.  Our
algorithms can process the largest dataset that has been used in
the  literature for exact DBSCAN, and outperform the
state-of-the-art distributed RP-DBSCAN algorithm~\cite{Song2018} by 18--577x.


The contributions of this paper are as follows.

\begin{enumerate}[label=(\textbf{\arabic*}),topsep=1pt,itemsep=0pt,parsep=0pt,leftmargin=15pt]
\item New parallel algorithms for 2D exact DBSCAN, and higher-dimensional  exact and approximate DBSCAN with  work bounds matching that of the best existing sequential algorithms, and polylogarithmic depth.
\item Highly-optimized implementations of our parallel DBSCAN algorithms.
  \item A comprehensive experimental evaluation showing that our algorithms achieve excellent parallel speedups over the best sequential algorithms and outperform existing parallel algorithms by up to orders of magnitude.
 \end{enumerate}

We have made our source code publicly available at {\color{blue}\url{https://github.com/wangyiqiu/dbscan}}.

\section{Preliminaries}\label{sec:prelims}

\begin{figure*}
\begin{center}
\includegraphics[width=\textwidth]{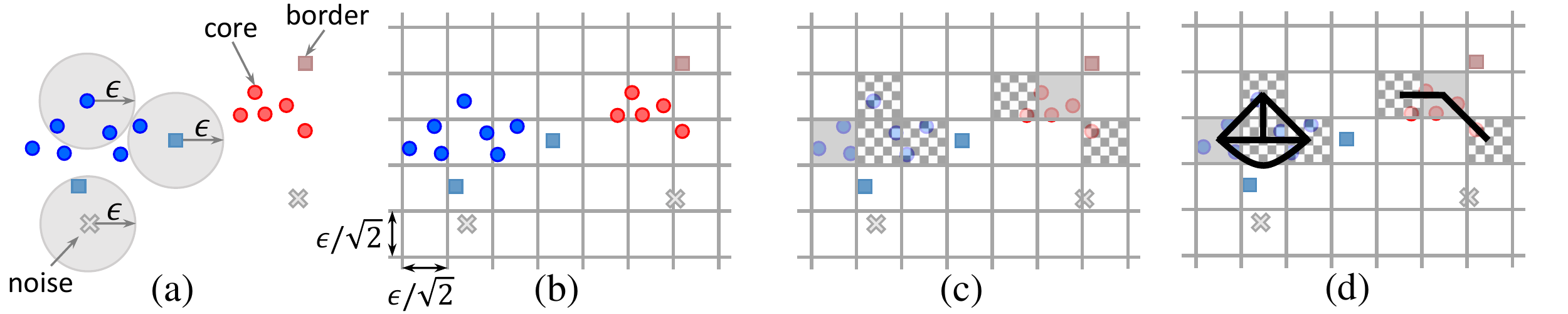}
\caption{\label{fig:overview} 
An example of DBSCAN and basic concepts in two dimensions.
Here we set $\minpts=3$ and $\epsilon$ as drawn.
In (a), the points are categorized into core points (circles) in two clusters (red and blue), border points (squares) that belong to the clusters, and noise points (crosses).
Using the grid method for cell construction, the algorithm
constructs cells with side length $\epsilon/\sqrt{2}$ (diagonal length $\epsilon$), as shown in~(b).
The cells with at least $\minpts$ points are marked as core cells (solid gray cells in (c)), while points in other cells try to check if they have $\minpts$ points within a distance of $\epsilon$.
If so, the associated cells are marked as core cells as well (checkered cells in (c)).
To construct the cell graph, we create an edge between two core cells if the closest pair of points from the two cells is
within a distance of $\epsilon$ (shown in (d)).
Each connected component in the cell graph is a unique cluster. Border points are assigned to clusters that they are within $\epsilon$ distance from.
}
\end{center}
\end{figure*}

\myparagraph{DBSCAN Definition}
The \defn{DBSCAN} (density-based spatial clustering of applications with noise)
problem takes as input $n$ points $\mathcal{P}=\{p_0,\ldots, p_{n-1}\}$, a distance function $d$, and two
  parameters $\epsilon$ and $\minpts$~\cite{Ester1996}.  A point $p$
  is a \defn{core point} if and only if $\left|\{p_i ~|~p_i\in
  \mathcal{P},d(p,p_i)\le \epsilon\}\right|\ge \minpts$.
  We denote the set
  of core points as $\mathcal{C}$.  DBSCAN computes and outputs
  subsets of $\mathcal{P}$, referred to as \defn{clusters}.  Each
  point in $\mathcal{C}$ is in exactly one cluster, and two points
  $p,q\in\mathcal{C}$ are in the same cluster if and only if there
  exists a list of points $\bar{p}_1=p, \bar{p}_2, \ldots,
  \bar{p}_{k-1}, \bar{p}_k=q$ in $\mathcal{C}$ such that
  $d(\bar{p}_{i-1},\bar{p}_{i})\le \epsilon$.  For all non-core points
  $p\in\mathcal{P}\setminus\mathcal{C}$, $p$ belongs to cluster $C_i$
  if $d(p,q)\le \epsilon$ for at least one point $q \in
  \mathcal{C}\cap C_i$. Note that a non-core point can belong to
  multiple clusters. A non-core point belonging to at least one
  cluster is called a \defn{border point} and a non-core point
  belonging to no clusters is called a \defn{noise point}. For a given
  set of points and parameters $\epsilon$ and $\minpts$, the clusters
  returned are unique.  Similar to many previous papers on parallel
  DBSCAN, we focus on the Euclidean distance metric in this paper.
  See Figure~\ref{fig:overview}(a) for an illustration of the DBSCAN problem.


Gan and Tao~\cite{GanT17} define the \defn{approximate DBSCAN}
problem, which in addition to the DBSCAN inputs, takes a parameter
$\rho$. The definition is the same as DBSCAN, except for the
connectivity rule among core points. In particular, core points within
a distance of $\epsilon$ are still connected, but core points within a
distance of $(\epsilon,\epsilon(1+\rho)]$ may or may not be
connected. Core points with distance greater than $\epsilon(1+\rho)$
are still not connected. Due to this relaxation, multiple valid
clusterings can be returned. The relaxation is what enables an
asymptotically faster algorithm to be designed.  A variation of this
problem was described by Chen et al.~\cite{chen2005geometric}.


  Existing algorithms as well as some of our new
  algorithms use subroutines for solving the \defn{bichromatic closest pair
    (BCP)} problem, which takes as input two sets of points $P_1$ and
  $P_2$, finds the closest pair of points $p_1$ and $p_2$ such that
  $p_1 \in P_1$ and $p_2 \in P_2$, and returns the pair and their distance.




\myparagraph{Computational Model} We use the work-depth
model~\cite{JaJa92,CLRS} to analyze the theoretical efficiency of parallel
algorithms.
The \defn{work} of an algorithm is
the number of operations used, similar to the time complexity in the
sequential RAM model.  The \defn{depth} is the length of the longest
sequence dependence.  
By Brent's scheduling theorem~\cite{Brent1974}, an algorithm with work
$W$ and depth $D$ has overall running time $W/P + D$, where $P$ is the
number of processors available.  In practice, the Cilk work-stealing
scheduler~\cite{Blumofe1999} can be used to obtain an expected running time
of $W/P+O(D)$.  A parallel algorithm is \defn{work-efficient} if its
work asymptotically matches that of the best serial algorithm for
the problem, which is important since in practice the $W/P$ term in
the running time often dominates.

\myparagraph{Parallel Primitives} We give an overview of the primitives used in
our new parallel algorithms, and show their work and depth bounds in Table~\ref{tab:prelim-table}.
We use implementations of these primitives from the Problem Based
Benchmark Suite (PBBS)~\cite{SBFG}, an open-source library.

\begin{table}[t]
      \setlength{\tabcolsep}{4pt}
  \centering
  \small
    \begin{tabular}{llll}
    \midrule
          & Work  & Depth & Reference \\
    \hline
    Prefix sum, Filter & $O(n)$  & $O(\log n)$ & \cite{JaJa92} \\
    Comparison sort & $O(n\log n)$ & $O(\log n)$ & \cite{JaJa92,Cole88} \\
    Integer sort & $O(n)$  & $O(\log n)$ & \cite{Vishkin10} \\
    Semisort & $O(n)^\dagger$ & $O(\log n)^*$ & \cite{gu15semisort} \\
    Merge & $O(n)$  & $O(\log n)$ & \cite{JaJa92} \\
    Hash table & $O(n)^*$ & $O(\log n)^*$ & \cite{Gil91a} \\

    2D Delaunay triangulation & $O(n\log n)^*$ & $O(\log n)^*$ & \cite{Reif1992}\\
    \midrule
    \end{tabular}%
    \caption{Work and depth bounds for parallel primitives.  $^\dagger$ indicates an expected bound and $^*$ indicates a high-probability bound. The integer sort is for a polylogarithmic key range.  The cost of the hash table is for $n$ insertions or queries.}
  \label{tab:prelim-table}%
\end{table}%


\defn{Prefix sum} takes as input an array $A$ of length $n$, and
returns the array $(0, A[0], A[0] + A[1], \ldots, \sum_{i=0}^{n-2}
A[i])$ as well as the overall sum, $\sum_{i=0}^{n-1} A[i]$.  Prefix
sum can be implemented by first adding the odd-indexed elements to the
even-indexed elements in parallel, recursively computing the prefix
sum for the even-indexed elements, and finally using the results on
the even-indexed elements to update the odd-indexed elements in parallel.
This algorithm takes $O(n)$ work and $O(\log n)$ depth~\cite{JaJa92}.

\defn{Filter} takes an array $A$ of size $n$ and a predicate $f$, and
returns a new array $A'$ containing elements $A[i]$ for which
$f(A[i])$ is true, in the same order as in $A$.  We first construct an
array $P$ of size $n$ with $P[i]=1$ if $f(A[i])$ is true and $P[i]= 0$
otherwise.  Then we compute the prefix sum of $P$. Finally, for each
element $A[i]$ where $f(A[i])$ is true, we write it to the output
array $A'$ at index $P[i]$ (i.e., $A'[P[i]]=A[i]$).  This algorithm
also takes $O(n)$ work and $O(\log n)$ depth~\cite{JaJa92}.

\defn{Comparison sorting} sorts
 $n$ elements based on a comparison function.
Parallel comparison sorting can be done in $O(n\log n)$ work
and $O(\log n)$ depth~\cite{JaJa92,Cole88}.  We use a cache-efficient
samplesort~\cite{BGS10} from PBBS which samples
$\sqrt{n}$ pivots on each level of recursion, partitions the keys
based on the pivots, and recurses on each partition in parallel.

We also use \defn{integer sorting}, which sorts integer keys
from a polylogarithmic range in $O(n)$ work and $O(\log n)$
depth~\cite{Vishkin10}. The algorithm partitions the keys into
sub-arrays and in parallel across all partitions, builds a histogram
on each partition serially.  It then uses a prefix sum on the counts
of each key per partition to determine unique offsets into a global
array for each partition. Finally, all partitions write their keys
into unique positions in the global array in parallel.

\defn{Semisort} takes as input $n$ key-value pairs, and groups pairs
with the same key together, but with no guarantee on the relative
ordering among pairs with different keys. Semisort also returns the
number of distinct groups.  We use the implementation
from~\cite{gu15semisort}, which is available in PBBS.  The algorithm
 first hashes the keys, and then selects a sample of the keys to
predict the frequency of each key. Based on the frequency of keys
 in the sample, we classify them into ``heavy keys'' and
``light keys'', and assign appropriately-sized arrays for each heavy
key and each range of light keys. Finally, we insert all keys into random
locations in the appropriate array and sort within the array.  This
algorithm takes $O(n)$ expected work and $O(\log n)$ depth with high probability.\footnote{
We say that a bound holds \defn{with high
  probability (w.h.p.)} on an input of size $n$ if it holds with
probability at least $1-1/n^c$ for a constant $c>0$.}

\defn{Merge} takes two sorted arrays, $A$ and $B$, and merges them into a single
sorted array. If the sum of the lengths of the inputs is $n$, this can
be done in $O(n)$ work and $O(\log n)$ depth~\cite{JaJa92}. The
algorithm takes equally spaced pivots $A$
and does a binary search for each pivot in $B$. Each
sub-array between pivots in $A$ has a corresponding sub-array between
the binary search results in $B$.  Then it repeats the above process
for each pair, except that equally spaced pivots are taken from
the sub-array from $B$ and binary searches are done in the sub-array
from $A$. This creates small subproblems each of which can be solved
using a serial merge, and the results are written to a unique range of
indices in the final output. All subproblems can be processed in
parallel.

For \defn{parallel hash tables},
we can perform $n$  insertions or queries taking $O(n)$ work and $O(\log n)$
depth w.h.p.~\cite{Gil91a}.
We use the non-deterministic
concurrent linear probing hash table from~\cite{ShunB14}, which
uses an atomic update to insert an element to an empty location in its
probe sequence, and continues probing if the update fails. 

The \defn{Delaunay triangulation} on a set of points in 2D contains
triangles among every triple of points $p_1$, $p_2$, and $p_3$ such
that there are no other points inside the circumcircle defined by
$p_1$, $p_2$, and $p_3$~\cite{BCKO}. Delaunay triangulation can be
computed in parallel in $O(n\log n)$ work and $O(\log n)$
depth w.h.p.\,\cite{Reif1992}.  We use the randomized incremental algorithm from PBBS, which
 inserts points in parallel into the
triangulation in rounds, such that the updates to the
triangulation in each round by different points do not conflict~\cite{BFGS12}.

\section{\highlight{DBSCAN Algorithm Overview}}\label{sec:algo-intro}

This section reviews the high-level structure of existing sequential DBSCAN
algorithms~\cite{Gunawan13,GanT17,BergGR17} as well as our new
parallel algorithms.  
The high-level structure is shown in
Algorithm~\ref{alg:dbscan-highlevel}, and an illustration of the key
concepts are shown in Figure~\ref{fig:overview}(b)-(d).

We place the points into
disjoint $d$-dimensional \defn{cells} with side-length  $\epsilon/\sqrt{d}$
based on their coordinates (Line 2 and Figure~\ref{fig:overview}(b)). The
cells have the property that all points inside a cell are
within a distance of $\epsilon$ from each other, and will belong to
the same cluster in the end.
Then on Line 3 and Figure~\ref{fig:overview}(c), we mark the core points.
On Line 4, we generate the clusters for core points as follows. We
create a graph containing one vertex per \defn{core cell} (a
cell containing at least one core point), and connect two
vertices if the closest pair of core points from the two cells is
within a distance of $\epsilon$. We refer to this graph as the \defn{cell graph}.
This step is illustrated in Figure~\ref{fig:overview}(d).
We then find the connected components
of the cell graph to assign cluster IDs to points in core cells. On Line 5,
we assign cluster IDs for border points.
Finally, we
return the cluster labels on Line 6.

\begin{algorithm}[t]
\small
  \caption{DBSCAN Algorithm}\label{alg:dbscan-highlevel}
  \begin{flushleft}
   \textbf{Input}: A set $\mathcal{P}$ of  points, $\epsilon$, and $\minpts$ \\
  \textbf{Output}: An array $\clusters$ of sets of cluster IDs for each  point
  \end{flushleft}

  \begin{algorithmic}[1]

    \Procedure{DBSCAN}{$\mathcal{P}$, $\epsilon$, $\minpts$}
    \State $\mathcal{G} := \Call{Cells}{\mathcal{P}$, $\epsilon}$
    \State $\coreFlags := \Call{MarkCore}{\mathcal{P}, \mathcal{G}, \epsilon, \minpts$}
    \State $\clusters := \Call{ClusterCore}{\mathcal{P}, \mathcal{G}, \coreFlags, \epsilon, \minpts}$ \label{alg:dbscan-highlevel:clustercore}
    \State $\Call{ClusterBorder}{\mathcal{P}, \mathcal{G}, \coreFlags, \clusters, \epsilon,\minpts$}
    \State \Return $\clusters$
    \EndProcedure
  \end{algorithmic}
\end{algorithm}

All of our algorithms share this common
structure.  In Section~\ref{sec:2d}, we introduce our 2D algorithms, and in
Section~\ref{sec:high-d}, we introduce our algorithms for higher
dimensions.  We analyze the complexity of our algorithms in
Section~\ref{sec:analysis}.

\section{2D DBSCAN Algorithms}\label{sec:2d}

This section presents our parallel algorithms for implementing each
line of Algorithm~\ref{alg:dbscan-highlevel} in two dimensions.  The cells can be
constructed either using a grid-based method or a box-based method,
which we describe in Sections~\ref{sec:grid} and~\ref{sec:box},
respectively. Section~\ref{sec:markcore} presents our algorithm for
marking core points. We present several methods for constructing the
cell graph in Section~\ref{sec:clustercore}. Finally,
Section~\ref{sec:clusterborder} describes our algorithm for clustering
border points.


\subsection{Grid Computation}\label{sec:grid}

In the grid-based method, the points are placed into disjoint
cells with side-length  $\epsilon/\sqrt{2}$
based on their coordinates, as done in the sequential algorithms by
Gunawan~\cite{Gunawan13} and de Berg et al.~\cite{BergGR17}.
A hash table is used to store only the non-empty cells, and a
serial algorithm simply inserts each point into the cell corresponding
to its coordinates.

\myparagraph{Parallelization} The challenge in parallelization is in
distributing the points to the cells in parallel while
maintaining work-efficiency.  While a comparison sort could be used to
sort points by their cell IDs, this approach requires $O(n\log n)$
work and is not work-efficient.
We observe that semisort  (see
Section~\ref{sec:prelims}) can be used to solve this problem
work-efficiently.  The key insight here is that we only need to group
together points in the same cell, and do not care about the relative
ordering of points within a cell or between different cells.
We apply a semisort on an array of length $n$ of key-value pairs, where each key is
the cell ID of a point and the value is the ID of the point.
This also returns the number of distinct
groups (non-empty cells).

We then create a parallel hash table of size equal to the number of
non-empty cells, where each entry stores the bounding box of a cell as
the key, and the number of points in the cell and a pointer to
the start of its points in the semisorted array as the value. We can
determine neighboring cells of a cell $g$ with arithmetic computation
based on $g$'s bounding box, and then look up each neighboring cell in
the hash table, which returns the information for that cell if it is
non-empty.

\begin{figure}
  \begin{center}
    \vspace{-8pt}
\includegraphics[width=\columnwidth]{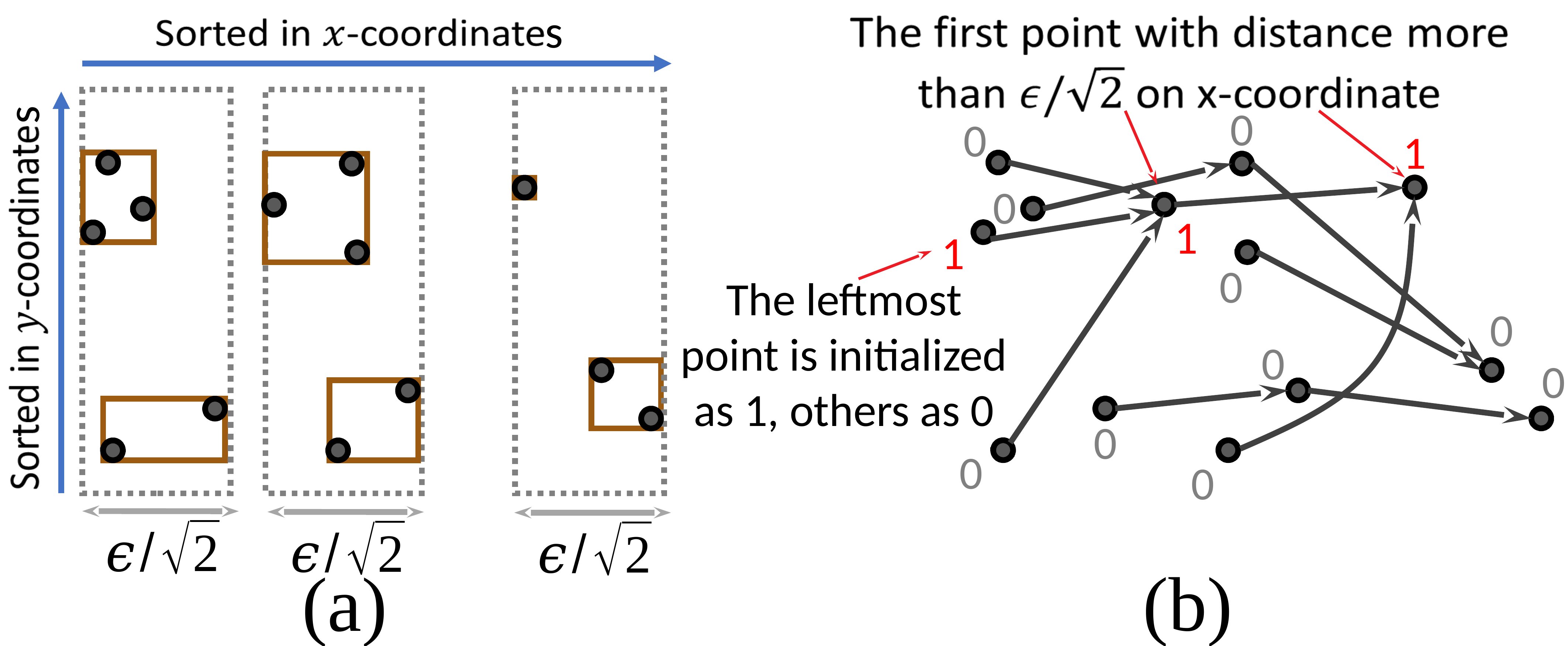}
\caption{\label{fig:box}
  Parallel box method construction.  In (a), the gray dashed rectangles correspond to strips and the brown solid rectangles correspond to box cells.  To compute the strips, we create a pointer from each point to the first point with an $x$-coordinate that is more than $\epsilon/\sqrt{2}$ larger. We initialize the leftmost point with a value of 1 and all other points with a value of 0. As shown in (b), after running pointer jumping, the points at the beginning of strips have values of 1 and all other points have values of 0.  We apply the same procedure in each strip on the $y$-coordinates to obtain the boxes.
}
\end{center}
\end{figure}

\subsection{Box Computation}\label{sec:box}
In the box-based method, we place the points into disjoint $2$-dimensional
bounding boxes with side-length at most $\epsilon/\sqrt{d}$, which are the cells.

Existing sequential solutions~\cite{Gunawan13,BergGR17} first sort all
points by $x$-coordinate, then scan through the points, grouping them
into strips of width $\epsilon/\sqrt{2}$ and starting a new strip when
a scanned point is further than distance $\epsilon/\sqrt{2}$ from the
beginning of the strip. It then repeats this process per strip in the
$y$-dimension to create cells of side-length at most
$\epsilon/\sqrt{2}$.  This step is shown in Figure~\ref{fig:box}(a).
Pointers to neighboring cells are stored per cell.  This is computed
for all cells in each $x$-dimensional strip $s$ by merging $s$ with
each of strips $s-2$, $s-1$, $s+1$, and $s+2$, as these are the only
strips that can contain cells with points within distance
$\epsilon$. For each merge, we compare the bounding boxes of the cells
in increasing $y$-coordinate, linking any two cells that may possibly
have points within $\epsilon$ distance.

\myparagraph{Parallelization} We now describe the method for assigning
points to strips, which is illustrated in Figure~\ref{fig:box}(b). Let
$p_x$ be the $x$-coordinate of point $p$.  We create a linked list
where each point is a node. The node for point $p$ stores a pointer to
the node for point $q$ (we call $q$ the \defn{parent} of $p$), where
$q$ is the point with the smallest $x$-coordinate such that
$p_x+\epsilon/\sqrt{2} < q_x$.
Each point can determine its parent in
$O(\log n)$ work and depth by binary searching the sorted list of
points.

We then assign a value of $1$ to the node with the smallest
$x$-coordinate, and $0$ to all other nodes.  We run a pointer jumping
routine on the linked list where on each round, every node passes its
value to its parent and updates its pointer to point to the parent of
its parent~\cite{JaJa92}. The procedure terminates when no more
pointers change in a round. In the end, every node with a value of $1$
will correspond to the point at the beginning of a strip, and all
nodes with a value of $0$ will belong to the strip for the closest
node to the left with a value of $1$. This gives the same strips as
the sequential algorithm, since all nodes marked $1$ will correspond
to the closest point farther than $\epsilon/\sqrt{2}$ from the point
of the previously marked node.
For merging to determine cells within distance $\epsilon$, we use the
parallel merging algorithm described in Section~\ref{sec:prelims}.


\subsection{Mark Core} \label{sec:markcore}

Illustrated in Figure~\ref{fig:overview}(c), the high-level idea in marking the core points is
as follows: first, if a cell contains at least $\minpts$ points
then all points in the cell are core points, as it is
guaranteed that all the points inside a cell will be within $\epsilon$
to any other point in the same cell; otherwise, each
point $p$ computes the number of points within its $\epsilon$-radius by
checking its distance to points in all \defn{neighboring cells}
(defined as cells that could possibly contain points within a distance
of $\epsilon$ to the current cell), and marking $p$ as a core point if
the number of such points is at least $\minpts$.
For a constant
dimension, only a constant number of neighboring cells
need to be checked.

\begin{algorithm}[!t]
  \small
  \caption{Parallel \textproc{MarkCore}}\label{alg:mark-core}
  
  \begin{algorithmic}[1]
    \Procedure{MarkCore}{$\mathcal{P}, \mathcal{G}, \epsilon, \minpts$}
    \State $\coreFlags := \{0,\ldots,0\}$ \Comment{Length $|\mathcal{P}|$ array} 
    \ParForEach {$g \in \mathcal{G}$}
    \If {$|g| \ge \minpts$} \Comment{$|g|$ is the number of points in $g$} 
      \ParForEach {$p$ in cell $g$}
      \State $\coreFlags[p] := 1$
      \EndFor
    \Else 
      \ParForEach {$p$ in cell $g$} \label{alg:mark-core:comp-start}
        \State $\arrcount := |g|$
        \ForEach {$h \in g.\Call{NeighborCells}{\epsilon}$}
        \State $\arrcount := \arrcount+\Call{RangeCount}{p, \epsilon, h}$
        \EndFor
        \If {$\arrcount \geq \minpts$}
        \State $\coreFlags[p] := 1$
        \EndIf
      \EndFor 
    \EndIf \label{alg:mark-core:comp-end} 
    \EndFor 
    \State \Return $\coreFlags$
    \EndProcedure
  \end{algorithmic}
\end{algorithm}

\myparagraph{Parallelization}
Our parallel algorithm for marking core points is shown in
Algorithm~\ref{alg:mark-core}.  We create an array $coreFlags$ of
length $n$ that marks which points are core points.  The array is
initialized to all 0's (Line~2).  We then loop through all cells in
parallel (Line~3). If a cell contains at least $\minpts$ points, we
mark all points in the cell as core points in parallel (Line 4--6).
Otherwise, we loop
through all points $p$ in the cell in parallel, and for each
neighboring cell $h$ we count the number of points within a distance
of $\epsilon$ to $p$, obtained using a \Call{RangeCount}{$p$,
$\epsilon$, $h$} query (Lines 8--11) that reports the number of
points in $h$ that are no more than $\epsilon$ distance from $p$.
The \Call{RangeCount}{$p$, $\epsilon$, $h$} query can be implemented
by comparing $p$ to all points in each neighboring cell $h$ in parallel, followed by a parallel prefix sum to obtain the number of points in the $\epsilon$-radius.
If the total count is at least $\minpts$, then $p$ is marked as a core point
(Lines 12--13).


\subsection{Cluster Core}\label{sec:clustercore}


The next step of the algorithm is to generate the cell graph (illustrated in Figure~\ref{fig:overview}(d)).
We present three approaches for determining the connectivity between cells in the cell graph. After obtaining the cell graph, we run a parallel connected components algorithm to cluster the core points. For the BCP-based approach, we describe an optimization that merges the BCP computation with the connected components computation using a lock-free union-find data structure.


\myparagraph{BCP-based Cell Graph} The problem of determining cell
connectivity can be solved by computing the BCP of core points between
two cells (recall the definition in Section~\ref{sec:prelims}), and
checking whether the distance is at most $\epsilon$.

Each cell runs a BCP computation with each of its neighboring cells to
check if they should be connected in the cell graph.  We execute all
BCP calls in parallel, and furthermore each BCP call can be
implemented naively in parallel by computing all pairwise distances in
parallel, writing them into an array containing point pairs and their
distances, and applying a prefix sum on the array to
obtain the BCP.  We apply two optimizations to speed up individual BCP
calls: (1) we first filter out points further than $\epsilon$ from the
other cell beforehand as done by Gan and Tao~\cite{GanT17}, and (2) we
iterate only until finding a pair of points with distance at most
$\epsilon$, at which point we abort the rest of the BCP computation,
and connect the two cells.
Filtering points can be done using a parallel filter.
To parallelize the early termination optimization,
 it is not efficient
to simply parallelize across all the point comparisons as this will
lead to a significant amount of wasted work. Instead, we divide the
points in each cell into fixed-sized blocks, and iterate over all
pairs of blocks. For each pair of blocks, we compute the distances of
all pairs of points between the two blocks in parallel by writing
their distances into an array.  We then take the minimum distance in
the array using a prefix sum, and return if the minimum is at most
$\epsilon$. This approach reduces the wasted work over the naive
parallelization, while still providing ample parallelism within each
pair of blocks.

\begin{figure}
\begin{center}
  \includegraphics[width=1\columnwidth]{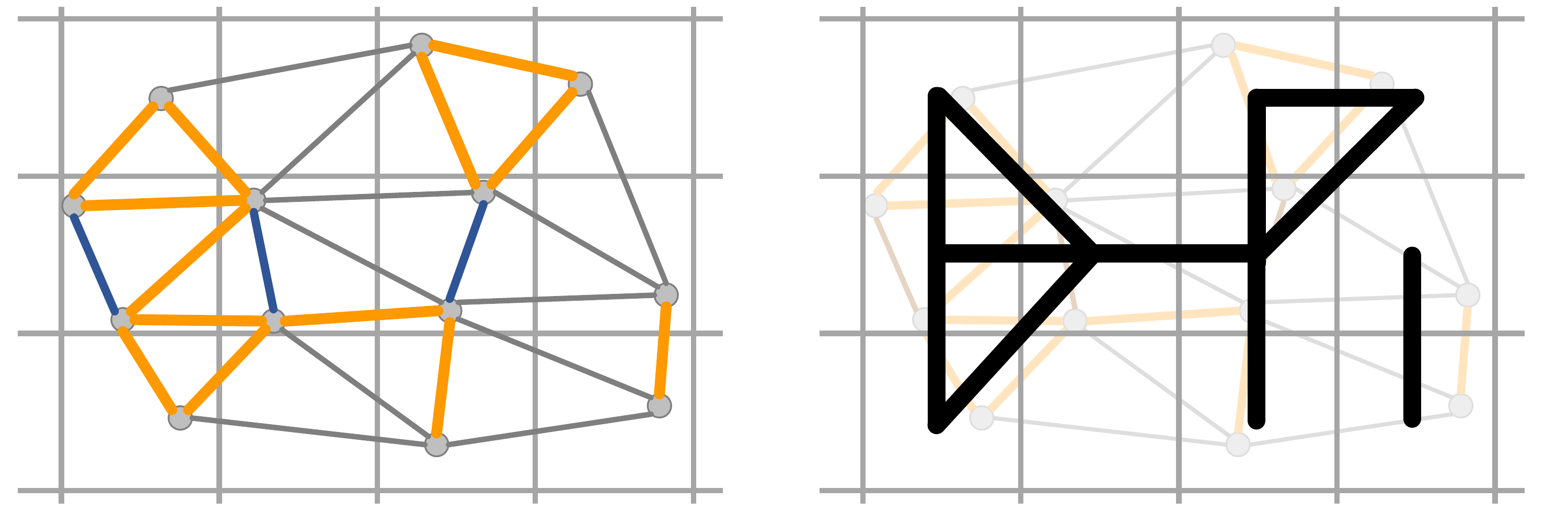}
\caption{\label{fig:delaunay} 
Using Delaunay triangulation (DT) to construct the cell graph in 2D.
\textbf{(Left)} We construct the DT for all core points, and an edge in the DT can either be inside a cell (dark blue), or across cells with length no more than $\epsilon$ (orange), or with length more than $\epsilon$ (gray).
\textbf{(Right)} An orange edge will add the associated edge in the cell graph, and in this example, there are two clusters.
}
\end{center}
\end{figure}

\myparagraph{Triangulation-based Cell Graph} In two dimensions,
Gunawan~\cite{Gunawan13} describes a special approach using Voronoi
diagrams. In particular, we can efficiently determine whether a core
cell should be connected to a neighboring cell by finding the nearest
core point from the neighboring cell to each of the core cell's core
points.  Gan and Tao~\cite{GanT17} and de Berg et al.~\cite{BergGR17}
show that a Delaunay triangulation can also be used to determine
connectivity in the cell graph. In particular, if there is an edge
in the Delaunay triangulation between two core cells with distance at
most $\epsilon$, then those two cells are connected.  This process is
illustrated in Figure~\ref{fig:delaunay}.  The proof of correctness is
described in~\cite{GanT17,BergGR17}.

To compute Delaunay triangulation or Voronoi diagram in parallel, Reif
and Sen present a parallel algorithm for constructing Voronoi diagrams
and Delaunay triangulations in two dimensions. We use the parallel
Delaunay triangulation implementation from PBBS~\cite{BFGS12,SBFG}, as
described in Section~\ref{sec:prelims}.

\begin{figure}
  \begin{center}
  \includegraphics[width=\columnwidth]{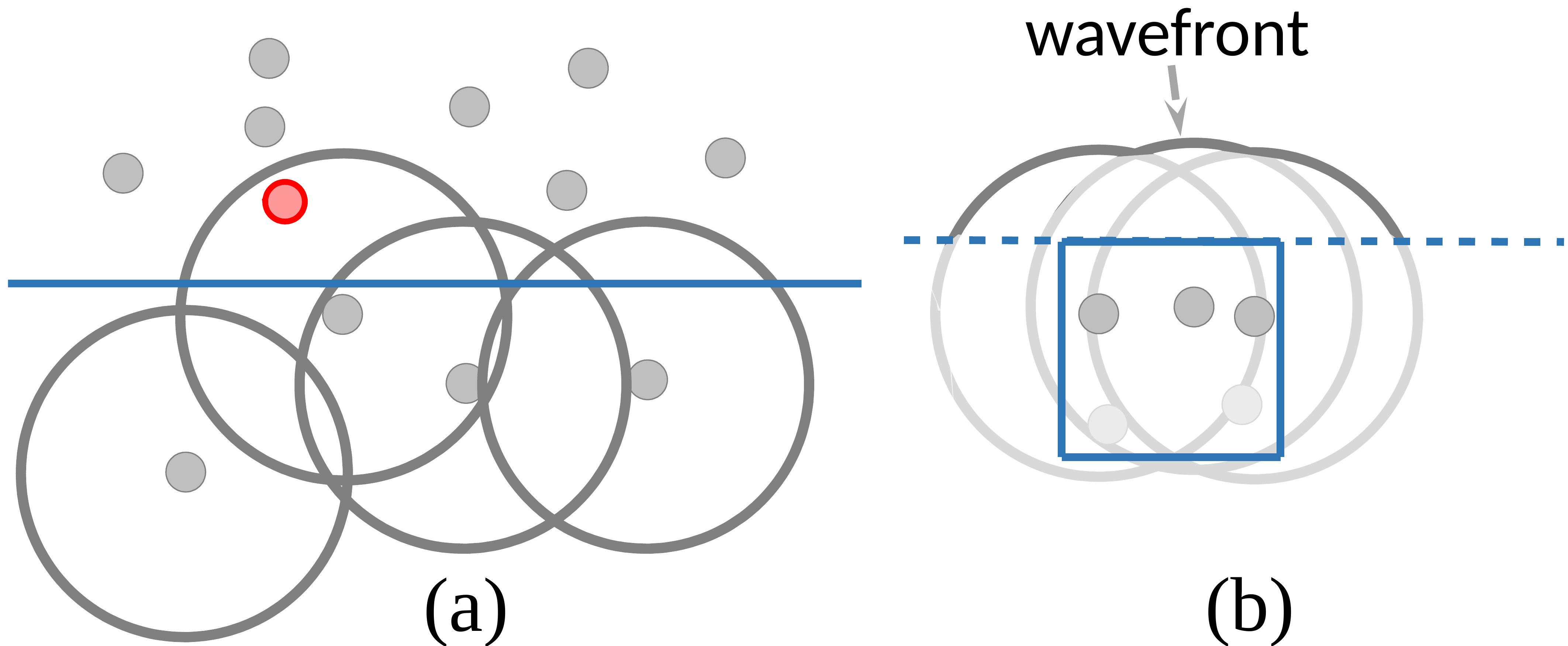}
\caption{\label{fig:usec} 
An example of the USEC with line separation problem.
In (a), the points are above the horizontal line while the circles are centered below the line.
In this case, the answer is ``yes'' since there is a point inside one of the circles.
In (b), we show how this problem relates to DBSCAN.
We generate the wavefront of the circles on the left and top borders of each cell, and check if core points in nearby cells are within the wavefront. In this example, we show the top wavefront.
}
\end{center}
\end{figure}

\myparagraph{Unit-spherical emptiness checking-based (USEC) Cell Graph} 
Gan and Tao~\cite{GanT17} (who attribute the idea to Bose et
al.~\cite{Bose2007}) describe an algorithm for solving the
unit-spherical emptiness checking (USEC) with line
separation problem when comparing two core cells to determine
whether they should be connected in the cell graph.

In the USEC with line separation problem, we are given a horizontal or
vertical line $\ell$ and would like to check if any of the
$\epsilon$-radius circles of points on one side of $\ell$ contain any
points on the other side of $\ell$.  The problem assumes that the
points on each side of $\ell$ are sorted by both $x$-coordinate and
$y$-coordinate.  This is illustrated in Figure~\ref{fig:usec}.

To apply the USEC with line separation problem to DBSCAN, the points
in each core cell are first sorted both by $x$-coordinate and by
$y$-coordinate (two copies are stored).
For each cell, we consider its top and left boundaries as $\ell$.
For each $\ell$, we generate
the union of $\epsilon$-radius circles centered around sorted points
in the cell (sorted by $x$-coordinate for a horizontal boundary and
$y$-coordinate for a vertical boundary) and keep the outermost arcs of
the union of circles lying on the other side of $\ell$, which is
called the wavefront.  This is illustrated in
Figure~\ref{fig:usec}(b).
For a cell connectivity query, we choose $\ell$ to be one of the
boundaries of the two cells that separates the two cells. Then we scan
the points of one cell in sorted order and check if any of them are
contained in the wavefront of the other cell.



Our algorithm assumes that all points are distinct.
Without loss of generality, assume that we are generating a wavefront
above a horizontal line. 
To generate the wavefront in parallel, we use
divide-and-conquer by constructing the wavefront for the left half of
the points and the right half of the points (in sorted order)
recursively in parallel.  Merging two wavefronts is more
challenging. 
The wavefronts are represented as balanced binary trees
supporting split and join operations~\cite{Akhremtsev2016}.
We merge two wavefronts by taking the top part of each wavefront and joining
them together. 
The top part of each wavefront can be obtained by checking where
the left and right wavefronts intersect, and then merging them.

We prove in Section~\ref{sec:usec_proof} of the Appendix
that the left and right wavefronts intersect at a unique point.
Denote the unique arc in the left wavefront that intersects with the right
wavefront as $A$. All the arcs to the right of $A$ in the left wavefront lie
under the right wavefront, so they will not form part of the
combined wavefront. On the other hand, all arcs to the left of $A$ in the left
wavefront forms the left half of the combined wavefront.
We find arc $A$ by doing an exponential search in the left wavefront starting from
the rightmost arc. For each arc $A'$ visited, we perform a binary search 
for $A'$ in the right wavefront to find its intersection with the right
wavefront. There are three possible results: (a) $A'$ lies completely under the right
wavefront (no intersection); (b) $A'$ intersects with the right wavefront;
and (c) $A'$ lies completely outside the right wavefront (no intersection).
If the result is (a),
we continue the exponential search; if the result is (b), we terminate the search
and join the two wavefronts at the intersection; and if
the result is (c), arc $A$ must lie between $A'$
and the previous arc visited in the exponential search, and so we continue with a
binary search inside that interval to find arc $A$.
After the entire wavefront is generated, we write it out to
an array by traversing the binary
tree in parallel.

We can perform a cell connectivity query in parallel by creating
sub-problems using pivots, similar to how parallel merge is
implemented (see Section~\ref{sec:prelims}).  Recall that we are
comparing the sorted points of one cell with the wavefront of the
other cell.  We take equally spaced arc intersections as pivots from
the wavefront, and use binary search to find where the pivot fits in
the sorted point set. Every set of arcs between two pivots corresponds
to a set of sorted points between two binary search results. Then we
repeat the procedure on each pair by taking equally spaced pivots in
the sorted point set and doing binary search of the pivot in the set
of arcs (which may split an arc into multiple pieces). This gives
subproblems containing a contiguous subset of the sorted points and a
contiguous subset of the (possibly split) arcs of the wavefront. Each
subproblem is solved using the sequential USEC with line separation
algorithm of~\cite{GanT17}. If any of the subproblems return
``yes'', then the answer to the original USEC with line separation
problem is ``yes'', and otherwise it is ``no''.

\begin{algorithm}[!t]
  \small
  \caption{Parallel \textproc{ClusterCore}}\label{alg:cluster-core}
  
  \begin{algorithmic}[1]
    \Procedure{ClusterCore}{$\mathcal{P}, \mathcal{G}, \codevar{coreFlags}, \epsilon, \minpts$}
    \State $\codevar{uf} := \Call{UnionFind()}{}$ \Comment{Initialize union-find structure}
    \State $\Call{SortBySize}{\mathcal{G}}$ \Comment{Sort by non-increasing order of size}
    \ParForEach {$\{g \in \mathcal{G} : g$ is core$\}$}
      \ForEach {$\{h \in g.\Call{NeighborCells}{\epsilon} : h$ is core$\}$}
        \If {$g > h$ and $\codevar{uf}.\Call{Find}{g} \neq \codevar{uf}.\Call{Find}{h}$}
        \If {$\Call{Connected}{g,h}$} \Comment{On core points only}
          \State $\codevar{uf}.\Call{Link}{g,h}$
        \EndIf
        \EndIf
      \EndFor
    \EndFor 

    \State $\clusters := \{-1,\ldots,-1\}$ \Comment{Length $|\mathcal{P}|$ array}
    \ParForEach {$\{g \in \mathcal{G} : g$ is core$\}$}
    \ParForEach {$\{p$ in cell $g$ : $\coreFlags[p] = 1\}$}

    \State $\clusters[p] := \codevar{uf}.\Call{Find}{g}$
    \EndFor
    \EndFor

    \State \Return $\clusters$
    \EndProcedure
  \end{algorithmic}
\end{algorithm}

\myparagraph{Reducing Cell Connectivity Queries}
We now present an optimization that merges the cell graph construction with the connected components computation using a parallel lock-free union-find
data structure to maintain the connected components on-the-fly. This technique is used in both the BCP approach and USEC approach for cell graph construction.
The pseudocode is shown in Algorithm~\ref{alg:cluster-core}.
The idea is to only run a cell connectivity query  between two cells if they are not yet in the same component (Line 6), which
 can reduce the total number of  connectivity queries.  For example, assume that
cells $a$, $b$, and $c$ belong to the same component.  After
connecting $a$ with $b$ and $b$ with $c$, we can avoid the
connectivity check between $a$ and $c$ by checking their respective
components in the union-find structure beforehand.  This optimization
was used by Gan and Tao~\cite{GanT17} in the sequential setting, and we
extend it to the parallel setting.
We also only check connectivity between two cells at most once by having the cell with
higher ID responsible for checking connectivity with the cell with a lower ID (Line 6).

When constructing the cell graph and checking connectivity, we  use
a heuristic to prioritize the cells based on the number of core points
in the cells, and start from the cells with more points, as shown on Line 3.
This is because cells with more points are more likely to have higher
connectivity, hence connecting the nearby cells together and pruning
their connectivity queries.
This optimization can be less efficient in parallel, since
a connectivity query could be executed before the corresponding query
that would have pruned it in the sequential execution.
To overcome this, we group the
cells into batches, and process each batch in parallel before moving
to the next batch. We refer to this new approach as \defn{bucketing}, and
show experimental results for it in Section~\ref{sec:exp}.

\subsection{Cluster Border}\label{sec:clusterborder}

To assign cluster IDs for border points. We check all points not yet
assigned a cluster ID, and for each point $p$, we check all of its
neighboring cells and add it to the clusters of all neighboring
cells with a core point within distance $\epsilon$ to $p$.

\begin{algorithm}[!t]
  \small
  \caption{Parallel \textproc{ClusterBorder}}\label{alg:cluster-border}
  \begin{algorithmic}[1]
    \Procedure{ClusterBorder}{$\mathcal{P},\!\mathcal{G},\!\coreFlags,\!\clusters,\!\epsilon,\!\minpts$}

    \ParForEach {$\{g \in \mathcal{G} : |g| < \minpts$\}} 
    \ParForEach {$\{p$ in cell $g : \coreFlags[p] = 0\}$} 


    \ForEach {$h \in g \cup g.\Call{NeighborCells}{\epsilon}$} 
    \ParForEach {$\{q$ in cell $h : \coreFlags[q] = 1\}\!$} \label{alg:cluster-border:parnbrpt} 
    \If{$d(p, q)$ $ \leq \epsilon$}
    \State $\clusters[p] := \clusters[p] \cup \clusters[q]$ \label{alg:cluster-border:paradd} \Comment{In parallel}
    \EndIf
    \EndFor
    \EndFor 
    \EndFor 
    \EndFor 
    \EndProcedure
  \end{algorithmic}
\end{algorithm}

\myparagraph{Parallelization}
Our algorithm is shown in
Algorithm~\ref{alg:cluster-border}.  We loop through all cells with
fewer than $\minpts$ points in parallel, and for each such cell we loop over
all of its non-core points $p$ in parallel (Lines 2--3). On Lines 4--7,
we check all core points
in the current cell $g$ and all neighboring cells, and if any are
within distance $\epsilon$ to $p$, we add their clusters to $p$'s
set of clusters (recall that border points can belong to multiple
clusters).

\section{Higher-dimensional Exact and Approximate DBSCAN}\label{sec:high-d}

The efficient exact and approximate algorithms for higher-dimensional
DBSCAN are also based on the high-level structure of
Algorithm~\ref{alg:dbscan-highlevel}, and are extensions of some of the techniques
for two-dimensional DBSCAN described in Section~\ref{sec:2d}.
They use the grid-based method for assigning points to cells (Section~\ref{sec:grid}).
Algorithms~\ref{alg:mark-core},~\ref{alg:cluster-core}, and~\ref{alg:cluster-border} are used for marking core points, clustering core points, and clustering border points, respectively.
However, we use two major optimizations on top of the 2D algorithms:
a $k$-d tree for finding neighboring cells and a quadtree for answering range counting queries. 

\subsection{Finding Neighboring Cells}\label{sec:neighbor-cell}


The number of possible neighboring cells grows exponentially with the
dimension $d$,
and so enumerating all
possible neighboring cells can be inefficient in practice for higher
dimensions (although still constant work in theory).
Therefore, instead of implementing \textproc{NeighborCells} by enumerating all possible neighboring cells,
we first insert all cells into a $k$-d tree~\cite{Bentley1975},
which enables us to perform range queries to obtain just the non-empty
neighboring cells.  The construction of our $k$-d tree is done
recursively, and all recursive calls for children nodes are executed
in parallel. We also sort the points at each level in parallel and
pass them to the appropriate child.   Queries do not modify the
$k$-d tree, and can all be performed in parallel.  Since
a cell needs to find its neighboring cells multiple times throughout
the algorithm, we cache the result on its first query to avoid
repeated computation.

\subsection{Range Counting}\label{sec:quadtree}

\begin{figure}
  \begin{center}
\includegraphics[width=\columnwidth]{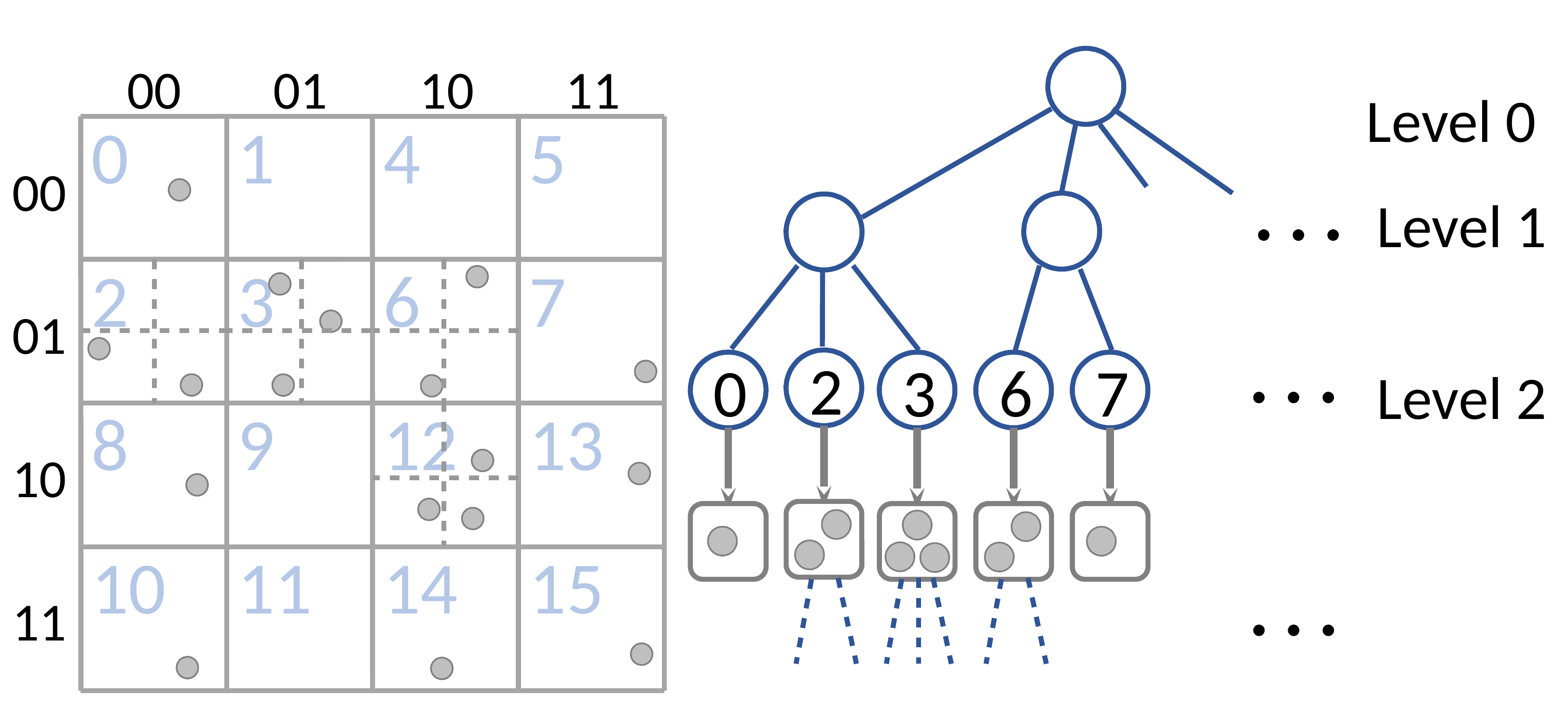}
\caption{A cell (left) and its corresponding
  quadtree data structure (right).\label{fig:quadtree} 
}
\end{center}
\end{figure}

While \textproc{RangeCount} queries can be implemented theoretically-efficiently in DBSCAN by checking all points in the target cell, there is a large overhead for doing so in practice.
In higher-dimensional DBSCAN, we construct a quadtree data structure for each cell to
answer \textproc{RangeCount} queries.
The structure of a quadtree is illustrated in
Figure~\ref{fig:quadtree}.  A cell of side-length
$\epsilon/\sqrt{d}$ is recursively divided into $2^d$ sub-cells of the
same size until the sub-cell becomes empty. 
This forms a tree where each
sub-cell is a node and its children are the up to $2^d$ non-empty
sub-cells that it divides into. Each node of the tree stores the
number of points contained in its corresponding sub-cell.
Queries do not modify the quadtrees and are therefore all executed in parallel. We now describe how to construct the quadtrees in parallel.

\myparagraph{Parallel Quadtree Construction}
The construction procedure recursively divides each cell into
sub-cells.  Each node of the tree has access to the points contained
in its sub-cell in a contiguous subarray that is part of a global
array (e.g., by storing a pointer to the start of its points in the
global array as well as the number of points that it represents).  We
use an integer sort on keys from the range $[0,\ldots,2^d-1]$ to sort
the points in the subarray based on which of the $2^d$ sub-cells it
belongs to. Now the points belonging to each of the child nodes are
contiguous, and we can recursively construct the up to $2^d$ non-empty
child nodes independently in parallel by passing in the appropriate
subarray.

To reduce construction time,
we set a
 threshold for the number of points in a sub-cell, below which the
 node becomes a leaf node. This reduces the height of the tree but
 makes leaf nodes larger.  In addition, we avoid unnecessary tree node
 traversal by ensuring that each tree node has at least two non-empty
 children: when processing a cell, we repeatedly divide the points
 until they fall into at least two different sub-cells.

\myparagraph{Range Counting in \textproc{MarkCore}}
\textproc{RangeCount} queries are used in marking core points  in
Algorithm~\ref{alg:mark-core}. For each cell, a quadtree
containing all of its points is constructed in parallel.  Then
the \Call{RangeCount}{$p$, $\epsilon$, $h$} query reports the number
of points in cell $h$ that are no more than $\epsilon$ distance from point
$p$.  Instead of naively looping through all points in $h$, we
initiate a traversal of the quadtree starting from cell $h$, and
recursively search all children whose sub-cell intersects with
the $\epsilon$-radius of $p$.  When reaching a leaf node on a query,
we explicitly count the number of points contained in the
$\epsilon$-radius of the query point.

\myparagraph{Exact DBSCAN}
For higher-dimensional exact DBSCAN, one of our implementations uses
 \textproc{RangeCount} queries when computing BCPs in
Algorithm~\ref{alg:cluster-core}.
For each core cell, we build a quadtree on its core points in parallel. 
Then for each core point $p$ in each core cell $g$, we issue a \textproc{RangeCount}
query to each of its neighboring core cells $h$ and connect $g$ and
$h$ in the cell graph if the range query returns a non-zero count of core points. Since we do not need to know the actual count, but only whether or not it is non-zero, our range query is optimized to terminate once such a result can be determined.
We combine this with the optimization of reducing cell connectivity queries described in Section~\ref{sec:clustercore}

\myparagraph{Approximate DBSCAN}
For approximate DBSCAN, the sequential algorithm of Gan and
Tao~\cite{GanT17} follows the high-level structure of
Algorithm~\ref{alg:dbscan-highlevel} using the grid-based cell
structure.  The only difference is in the cell graph construction,
which is done using approximate
\textproc{RangeCount} queries.  

In the quadtree for approximate \textproc{RangeCount}, each cell of
side-length $\epsilon/\sqrt{d}$ is still recursively divided into
$2^d$ sub-cells of the same size, but until either the sub-cell
becomes empty or has side-length at most $\epsilon\rho/\sqrt{d}$. The
tree has maximum depth $l=1+\lceil \log_2 1/\rho\rceil$.  We use a
modified version of our parallel quadtree construction method to
parallelize approximate DBSCAN.

An approximate \Call{RangeCount}{$p$, $\epsilon$, $h$, $\rho$} query
takes as input a point $p$, and returns an integer that is between
the number of points in the $\epsilon$-radius and the number of points
in the $\epsilon(1+\rho)$-radius of $p$ that are in $h$, (when using
approximate \textproc{RangeCount}, all relevant methods takes an
additional parameter $\rho$). If the answer is non-zero, then the core cell containing $p$ is connected to core cell $h$. 
Our query implementation starts a
traversal of the quadtree from $h$, and recursively searches all
children whose sub-cell intersects with the $\epsilon$-radius of
$p$. As done in exact DBSCAN, our query is optimized to terminate once a zero count or a non-zero count can be determined.
Once either a leaf node is reached or a node's sub-cell is
completely contained in the $\epsilon(1+\rho)$-radius of $p$, the
search on that path terminates.
Queries do not modify the quadtree and can all be executed in
parallel.


\section{Analysis} \label{sec:analysis}
This section analyzes the theoretical complexity of our algorithms, showing that they are work-efficient and have polylogarithmic depth.

\subsection{2D Algorithms}\label{sec:analysis-2d}

\myparagraph{Grid Computation}
In our parallel algorithm presented in Section~\ref{sec:grid}, creating
$n$ key-value pairs can be done in $O(n)$ work and $O(1)$ depth in a
data-parallel fashion.  Semisorting takes $O(n)$ expected work and
$O(\log n)$ depth w.h.p.  Constructing the hash table and inserting
non-empty cells into it takes $O(n)$ work and $O(\log n)$ depth
w.h.p. The overall cost of the parallel grid computation is therefore
$O(n)$ work in expectation and $O(\log n)$ depth w.h.p.

\myparagraph{Box Computation} 
The serial algorithm~\cite{Gunawan13,BergGR17} uses $O(n\log n)$ work, including sorting, scanning the points to assign them to strips and cells, and merging strips.
However, the span is $O(n)$ since in the worst case there can be  $O(n)$ strips.

Parallel comparison sorting takes
$O(n\log n)$ work and $O(\log n)$ depth. Therefore,
sorting the points by $x$-coordinate, and each strip by $y$-coordinate
can be done in $O(n\log n)$ work and $O(\log n)$ depth overall.
Parent finding using binary search for all points takes $O(n\log n)$ work
and $O(1)$ depth.
For pointer jumping, the longest path in the linked list halves on each round, and so the
algorithm terminates after $O(\log n)$ rounds. We do $O(n)$ work per
round, leading to an overall work of $O(n\log n)$. The depth is $O(1)$
per round, for a total of $O(\log n)$ overall. We repeat this process
for the points in each strip, but in the $y$-direction, and the work
and depth bounds are the same.
For assigning pointers to neighboring cells for each cell,
we use a parallel merging algorithm, which takes $O(n)$ work and
$O(\log n)$ depth. The pointers are stored in an array, accessible in constant work and depth.


\myparagraph{\textproc{MarkCore}} For cells with at least $\minpts$ points, we
spend $O(n)$ work overall marking their points as core points (Lines
4--6 of Algorithm~\ref{alg:mark-core}). All cells are processed in
parallel, and all points can be marked in parallel, giving $O(1)$
depth.

For all cells with fewer than $\minpts$ points, each point only needs
to execute a range count query on a constant number of neighboring
cells~\cite{Gunawan13,GanT17}.  \Call{RangeCount}{$p$, $\epsilon$,
  $h$} compares $p$ to all points in neighboring cell $h$ in parallel.
Across all queries, each cell will only be checked by $O(\minpts)$
many points, and so the overall work for range counting is
$O(n\cdot\minpts)$. Therefore, Lines 8--13 of
Algorithm~\ref{alg:mark-core} takes $O(n\cdot\minpts)$ work. All
points are processed in parallel, and there are a constant number of
\textproc{RangeCount} calls per point, each of which takes $O(\log n)$
depth for a parallel prefix sum to obtain the number of points in the
$\epsilon$-radius.  Therefore, the depth for range counting is $O(\log
n)$.



The work for looking up the neighbor cells is $O(n)$ and depth is $O(\log n)$
w.h.p. using the parallel hash table that stores the non-empty cells.
Therefore, parallel \textproc{MarkCore} takes $O(n\cdot\minpts)$ work and
 $O(\log n)$ depth w.h.p.

\myparagraph{Cell Graph Construction} Reif and Sen present a parallel
algorithm for constructing Voronoi diagrams and Delaunay
triangulations in two dimensions in $O(n\log n)$ work and $O(\log n)$
depth w.h.p.\,\cite{Reif1992}.  For the Voronoi diagram approach, each nearest
neighbor query can be answered in $O(\log n)$ work, which is used to check
whether two cells should be connected and can be applied in parallel.
Each cell will only execute a constant number of queries, and so the
overall complexity is $O(n \log n)$ work and $O(\log n)$ depth w.h.p. For
the Delaunay triangulation approach, we can simply apply a parallel
filter over all of the edges in the triangulation, keeping the edges
between different cells with distance at most $\epsilon$. The cost of
the filter is dominated by the cost of constructing the Delaunay
triangulation.  

For the USEC with line separation method, the sorted order of points
in each dimension for each cell can be generated in $O(n\log n)$ work
and $O(\log n)$ depth overall.  
To generate a wavefront on $n$ points, the exponential
search has $O(\log n)$ steps per level, and each step involves a binary search which takes $O(\log n)$ work
and depth.  Therefore, the the work and depth for the searches is $O(\log^2n)$ per level. Splitting and joining the binary trees to generate the new wavefront on each level
takes $O(\log n)$ work and depth~\cite{Akhremtsev2016}.
Thus, for the work, we obtain the recurrence
$W(n) = 2W(n/2)+O(\log^2 n)$, which solves to $O(n)$. Since we can solve
the recursive subproblems in parallel, for the depth, we obtain the recurrence
$D(n) = D(n/2)+O(\log^2 n)$, which solves to $O(\log^3 n)$.


Checking whether the sorted set of points from one cell intersects
with a wavefront from another cell can be done using an algorithm
similar to parallel merging, as described in
Section~\ref{sec:clustercore}.  In particular, for a wavefront of size
$m$ and sorted point set of size $s$, we pick $k=m/\log (m+s)$ equally
spaced pivots from the wavefront, and use binary search to split the
sorted point set into subsets of size $s_1,\ldots,s_{k+1}$ where
$\sum_{i=1}^{k+1}s_i = s$.  The binary searches take a total of
$O(k\log s)=O(m)$ work and $O(\log s)$ depth.  Then, for the $i$'th
pair forming a subproblem, we pick $j_i=s_i/\log(m+s)$ equally spaced
pivots from the $i$'th subset of points and perform a binary search into the
$i$'th subset of arcs of the wavefront to create more
subproblems. This takes a total of $\sum_{i=1}^{k+1}O(j_i\log m)=O(s)$
work and $O(\log m)$ depth. The subset of points and subset of arcs
for each subproblem are now all guaranteed to be of size $O(\log
(m+s))$.  In parallel across all subproblems, we run the serial USEC
with line separation algorithm of~\cite{GanT17}, which takes linear
work in the input size. Therefore, the work for this particular
instance of USEC with line separation is $O(m+s)$ and the depth is
$O(\log (m+s))$. All cell connectivity queries can be performed in
parallel, and so the total work is $O(n)$ and and depth is $O(\log
n)$.
Since the sequential algorithms for wavefront generation and
determining cell connectivity take linear work, our algorithm is
work-efficient. After generating each wavefront, we write it out to
an array by traversing the binary
tree in parallel, which takes linear work and logarithmic depth~\cite{Akhremtsev2016}.
Including the preprocessing step of sorting, our
parallel USEC with line separation problem for determining the
connectivity of core cells takes $O(n\log n)$ work and $O(\log^3n)$
depth. 

\myparagraph{Connected Components} After the cell graph that contains
$O(n)$ points and edges are constructed, we run connected components
on the cell graph.  This step can be done in parallel in $O(n)$
work and $O(\log n)$ depth w.h.p. using parallel connectivity
algorithms~\cite{Gazit1991,Halperin1994,Cole1996,Halperin2001,PettieR02}.

\myparagraph{\textproc{ClusterBorder}} Using a similar analysis as done for
marking core points, it can be shown that assigning cluster IDs to border
points takes $O(n\cdot \minpts)$ work
sequentially~\cite{Gunawan13,BergGR17}.  In parallel, since there are a
constant number of neighboring cells for each non-core point, and all
points in neighboring cells as well as all non-core points are checked
in parallel, the depth is $O(1)$ for the distance comparisons.
Looking up the neighboring cells can be done in $O(n)$
work and $O(\log n)$ depth w.h.p. using our parallel hash table.
Adding cluster IDs to border point's set of clusters, while removing
duplicates at the same time, can be done using parallel hashing in
linear work and $O(\log n)$ depth w.h.p.  The work is
$O(n\cdot\minpts)$ since we do not introduce any asymptotic work
overhead compared to the sequential algorithm.

Overall, we have the following theorem.

\begin{theorem}
For a constant value of $\minpts$, 2D Euclidean DBSCAN can be computed in $O(n \log n)$ work and  $O(\log n)$ depth w.h.p.
\end{theorem}

\subsection{Higher-dimensional Algorithm}

For $d\geq 3$ dimensions, the BCP problem can be solved either using
brute-force checking, which takes quadratic work, or using more
theoretically-efficient
algorithms that take sub-quadratic work~\cite{Agarwal1991,chazelle1985search,clarkson1988randomized}. This leads to a DBSCAN algorithm 
that takes $O((n\log n)^{4/3})$ expected work for $d=3$ and
$O(n^{2-(2/(\lceil d/2 \rceil + 1))+\delta})$ expected work for $d \ge
4$ where $\delta >0$ is any constant~\cite{GanT17}.
The
theoretically-efficient BCP algorithms seem too complicated to be
practical (we are not aware of any implementations of these
algorithms), and the actual implementation of~\cite{GanT17} does not
use them.  However, we believe that it is still theoretically
interesting to design a sub-quadratic work parallel BCP algorithm to
use in DBSCAN, which is the focus of this section.

The sub-quadratic work BCP algorithms are based on
constructing Delaunay triangulations (DT) in $d$ dimensions, which can
be used for nearest neighbor search. However, we cannot
afford to construct a DT on all the points, since a
$d$-dimensional DT contains up to $O(n^{\lceil d/2\rceil})$ simplices,
which is at least quadratic in the worst-case for $d\ge 3$.

The idea in the algorithm by Agarwal et al.\,\cite{Agarwal1991} is to
construct multiple DTs, each for a subset of the points, and a nearest
neighbor query then takes the closest neighbor among queries to all of
the DTs.  The data structure for nearest neighbor queries used by
Aggarwal et al. is based on the RPO (Randomized Post Office) tree by
Clarkson~\cite{clarkson1988randomized}.  The RPO tree contains $O(\log
n)$ levels where each node in the RPO tree corresponds to the DT of a
random subset of the points.  Parallel DT for a constant
dimension $d$ can be computed work-efficiently in expectation and in
$O(\log n\log^*n)$ depth w.h.p.\,\cite{BGSS2018}.
The children of each node can be determined by traversing the history
graph of the DT, which takes $O(\log n)$ work and depth.  The RPO tree
is constructed recursively for $O(\log n)$ levels, and so the overall
depth is $O(\log^2 n\log^*n)$ w.h.p.  A query traverses down a path in
the RPO tree, querying each DT along the path, which takes $O(\log^2
n)$ work and depth overall.  Using this data structure to solve BCP
gives a DBSCAN algorithm with $O(n^{2-(2/(\lceil d/2 \rceil +
  1))+\delta})$ expected work and $O(\log^2n\log^*n)$ depth w.h.p.
For $d=3$, an improved data structure by Agarwal et
al.\,\cite{Agarwal1990} can be used to improve the expected work to
$O((n\log n)^{4/3})$.  The data structure is also based on DT, and so
similar to before, we can parallelize the DT construction and obtain
the same depth bound.

The overall bounds are summarized in the following theorem.

\begin{theorem}
For a constant value of $\minpts$, Euclidean DBSCAN can be solved in
$O((n\log n)^{4/3})$ expected work for $d=3$ and $O(n^{2-(2/(\lceil
  d/2\rceil +1))+\delta})$ expected work for any constant $\delta>0$
for $d>3$, and polylogarithmic depth with high probability.
\end{theorem}

\subsection{Approximate Algorithm}\label{sec:approx-analysis}

The algorithms for grid construction, marking core points, connected components,
and clustering border points are the same as the exact algorithms, and so we
only analyze approximate cell graph construction in the approximate
algorithm based on the quadtree introduced in
Section~\ref{sec:quadtree}.  The quadtree has $l=1+\lceil \log_2
1/\rho\rceil$ levels and can be constructed in $O(n'l)$ work
sequentially for a cell with $n'$ points. A hash table is used to map
non-empty cells to their quadtrees, which takes $O(n)$ work w.h.p. to construct.
Using a fact from~\cite{Arya2000}, Gan and Tao show that the number of
nodes visited by a query is $O(1+(1/\rho)^{d-1})$. Therefore, for
constant $\rho$ and $d$, all of the quadtrees can be constructed in a
total of $O(n)$ work w.h.p., and queries can be answered in $O(1)$
expected work.

All of the quadtrees can be constructed in parallel.  To parallelize
the construction of a quadtree for a cell with $n'$ points, we sort
the points on each level in $O(n')$ work and $O(\log n')$ depth using
parallel integer sorting~\cite{Vishkin10}, since the keys are integers
in a constant range.  In total, this gives $O(n'l)$ work and $O(l\log
n')$ depth per quadtree. We use a parallel hash table to map non-empty cells to
their quadtrees, which takes $O(n)$ work and $O(\log n)$ depth
w.h.p. to construct.
To construct the cell graph, all core points issue a constant number
of queries to neighboring cells in parallel. The $O(n)$ hash table queries can be done in $O(n)$ work and $O(\log n)$ depth w.h.p. and thus
 cell graph construction has the same complexity. This gives the following theorem.

\begin{theorem}
For constant values of $\minpts$ and $\rho$, our approximate Euclidean
DBSCAN algorithm takes $O(n)$ work and $O(\log n)$ depth with
high probability.
\end{theorem}

\section{Experiments}\label{sec:exp}

This section presents  experiments  comparing our exact
and approximate  algorithms as well as existing algorithms.


\myparagraph{Datasets}
We use the synthetic seed spreader (SS) datasets produced by Gan and
Tao's generator~\cite{GanT17}. The generator produces points generated
by a random walk in a local neighborhood, but jumping to a random
location with some probability.  \defn{SS-simden} and \defn{SS-varden}
refer to the datasets with similar-density and variable-density
clusters, respectively.
We also use a synthetic dataset called
\defn{UniformFill} that contains points distributed uniformly at
random inside a bounding hypergrid with side length $\sqrt{n}$, where
$n$ is the total number of points. The points have double-precision floating point
values, but we scaled them to integers when testing Gan and Tao's
implementation, which requires integer coordinates.  We generated the
synthetic datasets with 10 million points (unless specified otherwise) for
dimensions $d=2,3,5,7$.

In addition, we use the following real-world datasets, which
contain points with double-precision floating point values.
\begin{enumerate}[label=(\textbf{\arabic*}),topsep=1pt,itemsep=0pt,parsep=0pt,leftmargin=15pt]
\item \defn{Household} \cite{UCI} is a 7-dimensional dataset with $2,049,280$
  points excluding the date-time information.
\item \defn{GeoLife}~\cite{Zheng2008} is a 3-dimensional dataset with $24,876,978$
points. This dataset contains user location data (longitude, latitude, altitude),
and its distribution is extremely skewed.
\item \defn{Cosmo50}~\cite{Kwon2010} is a 3-dimensional dataset with $321,065,547$
points. We extracted the $x$, $y$, and $z$ coordinate information to construct the 3-dimensional
dataset.
\item \defn{OpenStreetMap}~\cite{Haklay2008} is a 2-dimensional dataset with
$2,770,238,904$ points, containing GPS location data.
\item \defn{TeraClickLog}~\cite{TeraClickLog} is a 13-dimensional dataset with \\
$4,373,472,329$ points containing feature values and click feedback of online
  advertisements.
As far as we know, \textit{TeraClickLog} is the largest
dataset used in the literature for \emph{exact} DBSCAN.
\end{enumerate}

We performed a search on $\epsilon$ and $\minpts$ for the synthetic
datasets and chose the default parameters to be those that output a
correct clustering. For the \emph{SS} datasets, the default parameters that we use are
similar to those found by Gan and Tao~\cite{GanT17}.  For ease of
comparison, the default parameters for \textit{Household} are the same
as Gan and Tao~\cite{GanT17} and the default parameters for
\textit{GeoLife}, \textit{Cosmo50}, \textit{OpenStreetMap}, and
\textit{TeraClickLog} are same as RP-DBSCAN~\cite{Song2018}.  For
approximate DBSCAN, we set $\rho=0.01$, unless specified otherwise.

\myparagraph{Testing Environment}
We perform all of our experiments on Amazon EC2 machines. We use a
c5.18xlarge machine for testing of all datasets other than
\textit{Cosmo50}, \textit{OpenStreetMap}, and \textit{TeraClickLog}.
The c5.18xlarge machine has 2 $\times$ Intel Xeon Platinum 8124M
(3.00GHz) CPUs for a total for a total of 36 two-way hyper-threaded cores, and
144 GB of RAM.  We use a r5.24xlarge machine for the three larger
datasets just mentioned.  The r5.24xlarge machine has 2 $\times$ Intel
Xeon Platinum 8175M (2.50 GHz) CPUs for a total of 48 two-way hyper-threaded
cores, and 768 GB of RAM.
By default, we use all of the
cores with hyper-threading on each machine.
We compile our programs with the \texttt{g++} compiler (version 7.4) with the \texttt{-O3} flag, and use Cilk Plus for parallelism~\cite{cilkplus}.

\subsection{Algorithms Tested}

We implement the different methods for marking core points and BCP computation in
exact and approximate DBSCAN for $d\geq 3$, and present results for the fastest versions, which are described below.

\begin{itemize}[topsep=1pt,itemsep=0pt,parsep=0pt,leftmargin=10pt]
\item \defn{our-exact}: This exact implementation implements the \textproc{RangeCount} query in marking core points by scanning through all points in the neighboring cell in parallel described in Section~\ref{sec:markcore}. For determining connectivity in the cell graph, it uses the BCP method described in Section~\ref{sec:clustercore}.
\item \defn{our-exact-qt}: This exact implementation implements the \textproc{RangeCount} query supported by the quadtree described in Section~\ref{sec:quadtree}. For determining connectivity in the cell graph, it uses the BCP method described in Section~\ref{sec:clustercore}.
\item \defn{our-approx}: This approximate implementation implements the \textproc{RangeCount} query in marking core points by scanning through all points in the neighboring cell in parallel, and uses the quadtree for approximate \textproc{RangeCount} queries in cell graph construction described in Section~\ref{sec:quadtree}.
\item \defn{our-approx-qt}: This approximate implementation is the same as \emph{our-approx} except that it uses the \textproc{RangeCount} query supported by the quadtree described in Section~\ref{sec:quadtree} for marking core points.
\end{itemize}

We append the \defn{-bucketing} suffix to the names of these implementations
when using
the bucketing optimization described in
Section~\ref{sec:clustercore}.



For $d=2$, we have six implementations that differ in whether they
use the grid or the box method to construct cells and whether they use
BCP, Delaunay triangulation, or USEC with line separation to construct
the cell graph. We refer to these as \defn{our-2d-grid-bcp},
\defn{our-2d-grid-usec}, \defn{our-2d-grid-delaunay},
  \defn{our-2d-box-bcp}, \defn{our-2d-box-usec}, and
  \defn{our-2d-box-delaunay}.

We note that our exact algorithms return the same answer as the standard DBSCAN definition, and our approximate algorithms return answers that satisfy Gan and Tao's approximate DBSCAN definition (see Section~\ref{sec:prelims}).


We compare with the following implementations:
 \begin{itemize}[topsep=1pt,itemsep=0pt,parsep=0pt,leftmargin=10pt]
\item \defn{Gan\&Tao-v2}~\cite{GanT17} is the state-of-the-art serial implementation
  for both exact and approximate DBSCAN.
  \textit{Gan\&Tao-v2} only accepts integer values between
  $0$ and $100,000$, and so when running their code we scaled the
  datasets up into this integer range and scaled up the $\epsilon$
  value accordingly to achieve a consistent clustering output with
  other methods.
\item \defn{pdsdbscan}~\cite{Patwary2013} is the implementation of the
  parallel disjoint-set exact DBSCAN by Patwary et al. compiled with
  OpenMP.

\item \defn{hpdbscan}~\cite{Gotz2015} is the implementation of
  parallel exact DBSCAN by Gotz et al. compiled with OpenMP. We
  modified the source code to remove the file output code.

\item \defn{rpdbscan}~\cite{Song2018} is the state-of-the-art
  distributed implementation for DBSCAN using Apache Spark. We note that their
  variant does not return the same result as DBSCAN.
  We tested \textit{rpdbscan} on the same machine that we used, and also report the timings 
  in~\cite{Song2018}, which were obtained using at least as many cores as our largest
  machine.

\end{itemize}

\begin{figure*}[th]
\begin{center}
\includegraphics[width=1\textwidth]{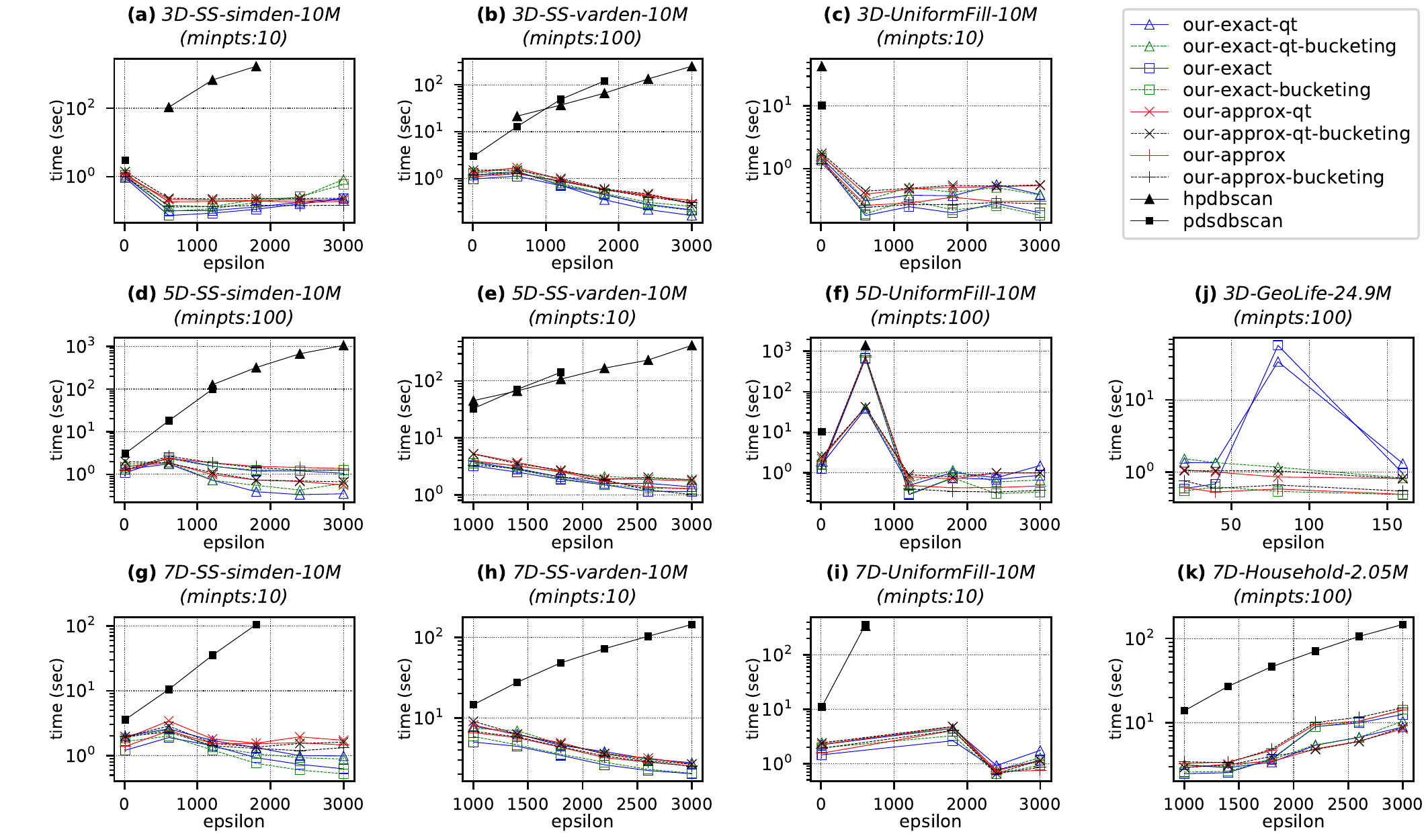}
\caption{\label{fig:ndeps} Running time vs. $\epsilon$ on 36 cores with hyper-threading. The $y$-axes are in log-scale.}
\end{center}
\end{figure*}

\subsection{Experiments for $d \geq 3$}

We first evaluate the performance of the different algorithms for
$d\geq 3$.  In the following plots, data points that did not finish within
an hour are not shown.

\myparagraph{Influence of $\mathbf{\epsilon}$ on Parallel Running
  Time} In this experiment, we fix the default value of $\minpts$ corresponding to the
correct clustering, and vary $\epsilon$ within a range centered around
the default $\epsilon$ value.
Figure~\ref{fig:ndeps} shows the parallel running time vs. $\epsilon$
for the different implementations.
In general, both \emph{pdsdbscan} and \emph{hpdbscan} becomes slower
with increasing $\epsilon$. This is because they use pointwise range queries, which get more expensive with larger $\epsilon$.
Our methods tend to improve with
increasing $\epsilon$ because there are fewer cells leading to a smaller cell graph, which speeds up computations on the graph.
Our implementations
significantly outperform \emph{pdsdbscan} and
\emph{hpdbscan} on all of the data points.

We observe a spike in plot Figure~\ref{fig:ndeps}(f) when
$\epsilon=608$. The implementations that mark core points by scanning
through all points in neighboring cells spend a significant amount of
time in that phase; in comparison, the quadtree versions perform
better because of their more optimized range counting. There is also a
spike in Figure~\ref{fig:ndeps}(j) when $\epsilon=80$.  Our exact
implementation spends a significant amount of time in cell graph
construction. This is because the \textit{GeoLife} dataset is heavily
skewed, certain cells could contain significantly more points. When
many cell connectivity queries involve these cells, the quadratic
nature using the BCP approach in \emph{our-exact} makes the cost of
queries expensive. On the contrary, methods using the quadtree for
cell graph construction (\emph{our-exact-qt}, \emph{our-approx-qt},
and \emph{our-approx}) tend to have consistent performance across the
$\epsilon$ values. For the spike in Figure~\ref{fig:ndeps}(j), it is interesting to see that the bucketing
implementations, \emph{our-exact-qt-bucketing} and
\emph{our-exact-bucketing}, are significantly faster than
\emph{our-exact-qt} and \emph{our-exact} because many of the expensive
connectivity queries are pruned.

\begin{figure*}
\begin{center}
\includegraphics[width=1\textwidth]{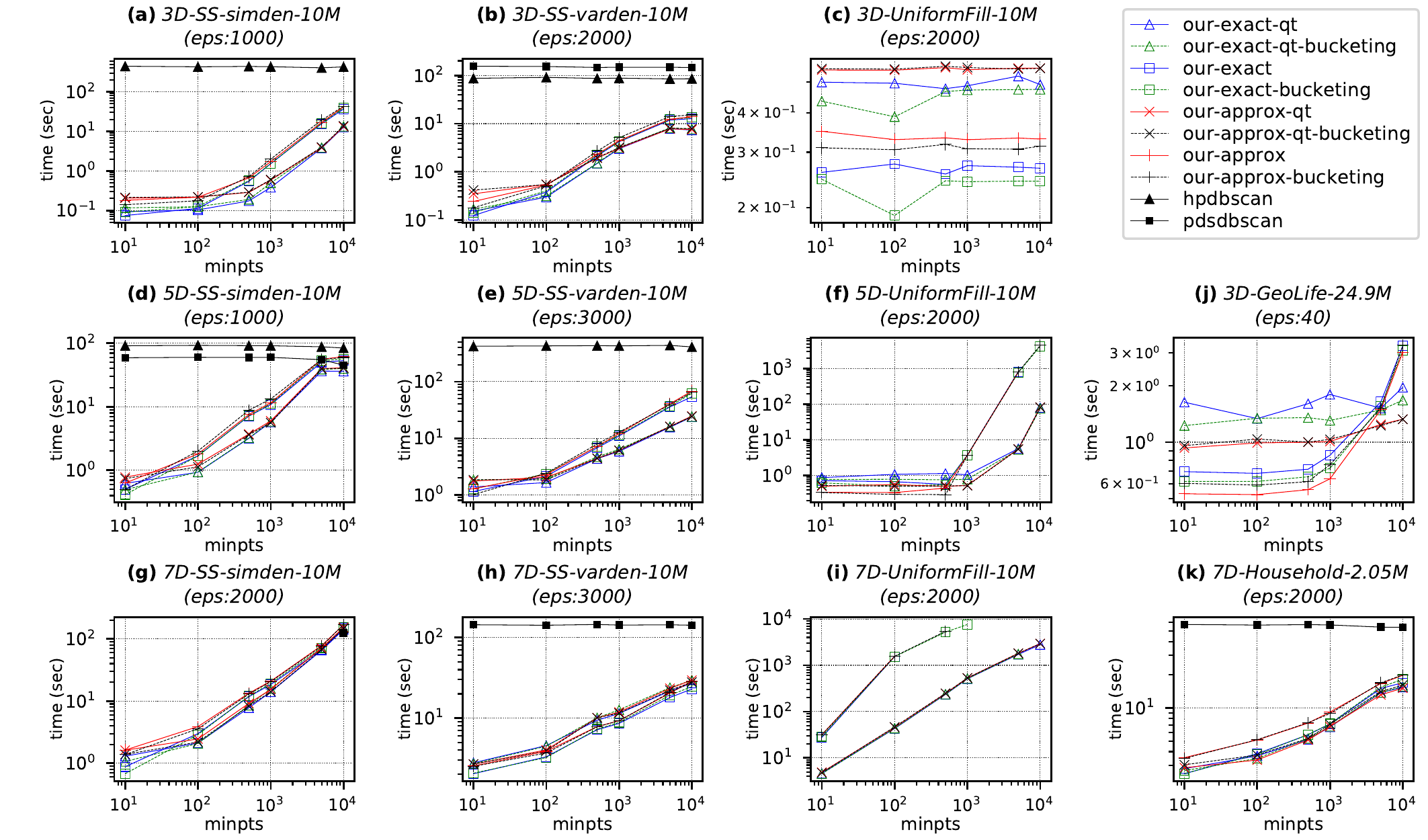}
\caption{Running time vs. $\minpts$ on 36 cores with hyper-threading. The $y$-axes are in log-scale.}
\label{fig:ndminpts}
\end{center}
\end{figure*}

\myparagraph{Influence of $\minpts$ on Parallel Running Time} In this
experiment, we fix the default value of $\epsilon$ for a dataset and
vary $\minpts$ over a range from $10$ to $10,000$.
Figure~\ref{fig:ndminpts} shows that our implementations have an
increasing trend in running time as $\minpts$ increases in most cases.
This is consistent with our analysis in Section~\ref{sec:analysis-2d}
that the overall work for marking core points is $O(n\cdot\minpts)$.
In contrast, $\minpts$ does not have much impact on the performance of
  \emph{hpdbscan} and \emph{pdsdbscan} because their range queries,
  which dominate the total running times, do not depend on $\minpts$.
Our implementations outperform \emph{hpdbscan} and \emph{pdsdbscan}
for almost all values of $\minpts$.  Figures~\ref{fig:ndminpts}(d)
and~\ref{fig:ndminpts}(g) suggests that \textit{hpdbscan} can surpass
our performance for certain datasets when $\minpts=10,000$. However,
as suggested by Schubert et al.\,\cite{Schubert2017}, the $\minpts$
value used in practice is usually much smaller, and based on our
observation, a $\minpts$ value of at most 100 usually gives the
correct clusters.

\begin{figure*}
\begin{center}
\includegraphics[width=1\textwidth]{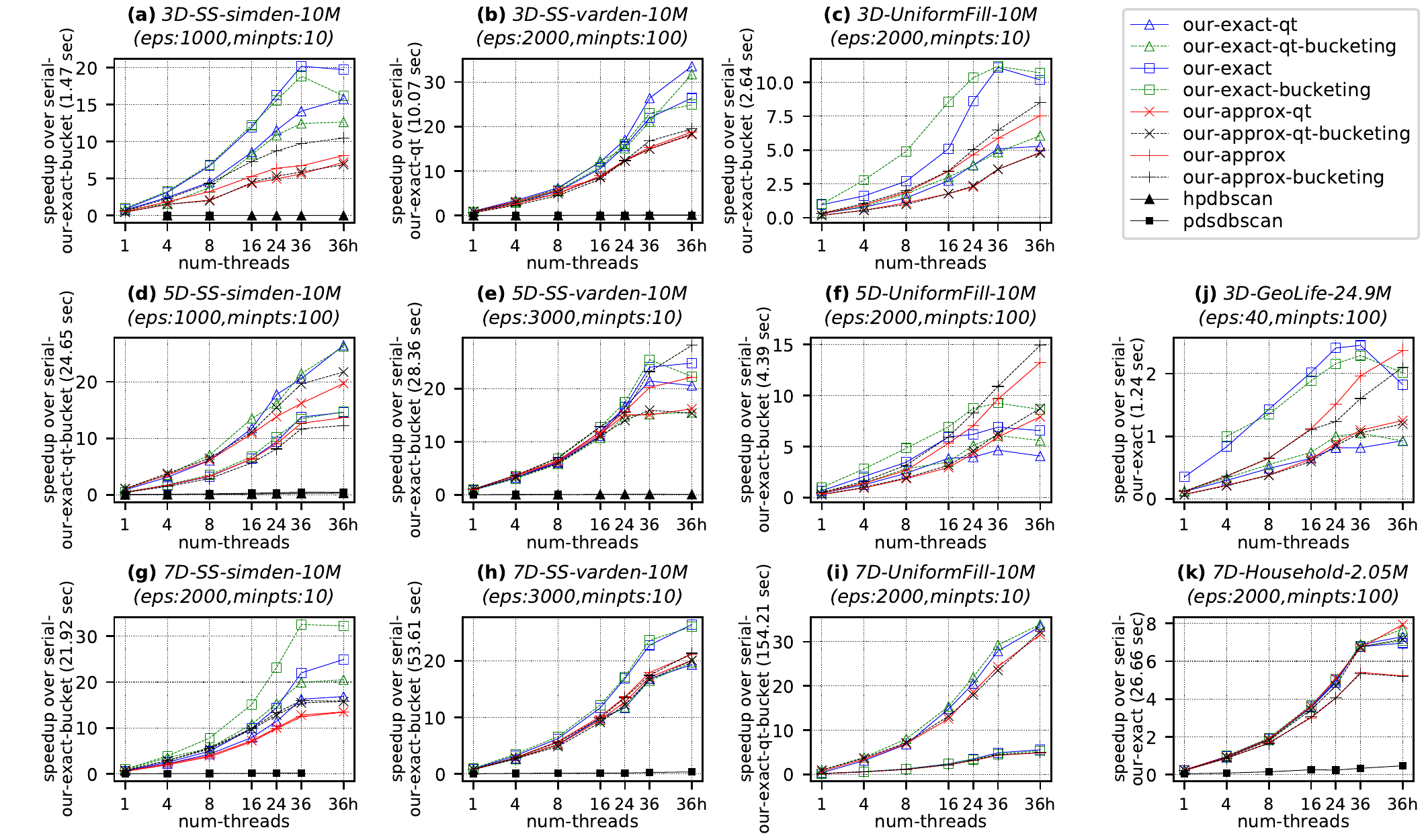}
\caption{Speedup of implementations over the \emph{best} serial baselines vs. thread count. The best serial baseline and its running time for each dataset is shown on the $y$-axis label. ``36h''
    on the $x$-axes refers to 36 cores with hyper-threading.
}
\label{fig:ndspeedup}
\end{center}
\end{figure*}

\begin{figure}
\begin{center}
\includegraphics[width=0.45\textwidth]{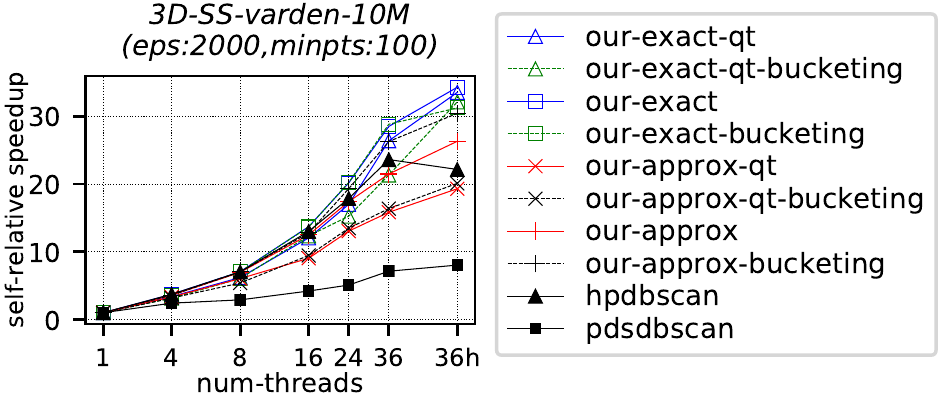}
\caption{Self-relative speedup of implementations vs. thread count.
    ``36h'' on the $x$-axis refers to 36 cores with hyper-threading.
}
\label{fig:ndspeedupself}
\end{center}
\end{figure}

\myparagraph{Parallel Speedup}
To the best of our knowledge, \textit{Gan\&Tao-v2} is the fastest existing serial implementation
both for exact and approximate DBSCAN. However, we find that across all of our datasets, our
serial implementations are faster than
theirs by an average of
5.18x and 1.52x for exact DBSCAN and approximate DBSCAN, respectively.
In Figure~\ref{fig:ndspeedup}, we compare the speedup of the parallel implementations
under different thread counts over the best serial baselines for each dataset and choice of parameters. We also show the self-relative speedups for one dataset in Figure~\ref{fig:ndspeedupself} and note that the trends are similar on other datasets.
For these experiments, we use parameters that generate the correct clusters.
Our implementations obtain very good speedups on most datasets, achieving speedups of 5--33x (16x on average) over the best serial baselines.
Additionally, the self-relative speedups of our exact and approximate methods are 2--89x (24x on average) and 14-44x (24x on average), respectively.
Although \emph{hpdbscan} and \emph{pdsdbscan} achieve good self-relative speedup (22--31x and 7--20x, respectively),
they fail to outperform the serial implementation on most of the datasets.
Compared to \emph{hpdbscan} and \emph{pdsdbscan}, we are faster by up to orders of magnitude (16--6102x).


Our speedup on the GeoLife dataset (Figure~\ref{fig:ndspeedup}(j)) is
low due to the high skewness of cell connectivity queries caused by
the skewed point distribution, however the parallel running time is
reasonable (less than 1 second). In contrast, \emph{hpdbscan} and
\emph{pdsdbscan} did not terminate within an hour.

The bucketing heuristic achieved the best
  parallel performance for several of the datasets
  (Figures~\ref{fig:ndeps}(f), (g), and (j); Figures~\ref{fig:ndminpts}(c) and (j); and Figures~\ref{fig:ndspeedup}(c), (f), (g), and (j)).
In general, the bucketing heuristic  greatly reduces the number of
connectivity queries during cell graph construction, but in some cases it can reduce parallelism and/or increase overhead due to sorting.
We also observe a similar trend on all methods where bucketing is applied.



We also implemented our own parallel baseline based on the original
DBSCAN algorithm~\cite{Ester1996}. We use a parallel $k$-d tree, and
all points perform queries in parallel to find all neighbors in their
$\epsilon$-radius to check if they should be a core
point. However, the baseline was over 10x slower than
our fastest parallel implementation for datasets with the correct
parameters, and hence we do not show it in the plots.

\begin{table*}[ht]
  \small
    \setlength{\tabcolsep}{4pt}
  \centering
\begin{tabular}{|c|c|c|c|c|c|c|c|c|c|c|c|c|c|c|c|c|}
\hline
          & \multicolumn{4}{c|}{GeoLife}  & \multicolumn{4}{c|}{Cosmo50} & \multicolumn{4}{c|}{OpenStreetMap} & \multicolumn{4}{c|}{TeraClickLog} \\ \hline
$\epsilon$       & 20    & 40    & 80    & 160   & 0.01  & 0.02  & 0.04  & 0.08 & 0.01    & 0.02   & 0.04   & 0.08   & 1500    & 3000   & 6000  & 12000  \\ \hline
\textit{our-exact} & 0.541 & 0.617 & 0.535 & 0.482 & 41.8  & 5.51  & 4.69  & 3.03 & 41.4    & 43.2   & 40     & 44.5   & 26.8    & 26.9   & 27.0  & 27.6   \\ \hline
\textit{rpdbscan (our machine)} & 29.13    & 27.92    & 32.04    & 27.81    & 3750   & 562.0   & 576.9   & 672.6  & --    & --   & --   & --    & --   & --   & --  & --   \\ \hline
\textit{rpdbscan (\cite{Song2018})} & 36    & 33    & 28    & 27    & 960   & 504   & 438   & 432  & 3000    & 1720   & 1200   & 840    & 15480   & 7200   & 3540  & 1680   \\ \hline
\end{tabular}
\caption{Parallel running times (seconds) for \emph{our-exact} and \textit{rpdbscan}. The value of $\minpts$ is set to $100$. \textit{GeoLife} was run on the 36 core machine and the other datasets were run on the 48 core machine. For \textit{rpdbscan}, we omit timings for experiments that encountered exceptions or did not complete within 1 hour. We also include the distributed running times reported in~\cite{Song2018} that used as many cores as our machines.  } \label{tab:largescale}
\end{table*}

\begin{figure}
  \begin{center}
\includegraphics[width=0.5\textwidth]{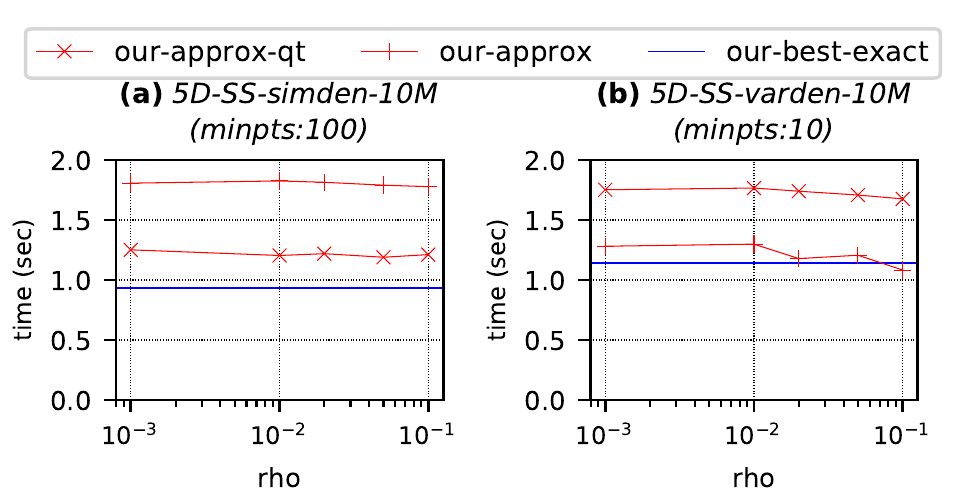}
\caption{Running time vs. $\rho$ on 36 cores with hyper-threading.}
 \label{fig:ndrhosmall}
\end{center}
\end{figure}

\myparagraph{Influence of $\mathbf\rho$ on Parallel Running Time}
Figure~\ref{fig:ndrhosmall} shows the effect of varying $\rho$ for our
two approximate DBSCAN implementations. We also show our best exact
method as a baseline. We only show plots for two datasets as the trend
was similar in other datasets.  We observe a small decrease in running
time as $\rho$ increases, but find that the approximate methods are
still mostly slower than the best exact method. On average, for the
parameters corresponding to correct clustering, we find that our best
exact method is 1.24x and 1.53x faster than our best approximate
method when running in parallel and serially, respectively; this can
also be seen in Figure~\ref{fig:ndspeedup}. Schubert et
al.\,\cite{Schubert2017} also found exact DBSCAN to be faster than
approximate DBSCAN for appropriately-chosen parameters, which is
consistent with our observation.

\myparagraph{Large-scale Datasets} In Table~\ref{tab:largescale}, we
show the running times of \textit{our-exact} on large-scale
datasets. We compare with the reported numbers for the
state-of-the-art distributed implementation \textit{rpdbscan}, which
use 48 cores distributed across 12 machines~\cite{Song2018}, as well
as numbers for \emph{rpdbscan} on our machines.  The purpose of this
experiment is to show that we are able to efficiently process large
datasets using just a multicore machine.  \textit{GeoLife} was run on
the 36 core machine whereas others were run on the 48 core machine due
to their larger memory footprint.
We see that \textit{our-exact} achieves a 18--577x speedup
over \textit{rpdbscan} using the same or a fewer number of cores. We believe that this speedup is due to lower communication costs in shared-memory as well as a better algorithm.
Even though \textit{TeraClickLog} is
significantly larger than the other datasets, our running times are
not proportionally larger. This is because for the parameters chosen
by~\cite{Song2018}, all points fall into one cell. Therefore, in our
implementation all points are core points and are trivially placed
into the only cluster. In contrast, \emph{rpdbscan} incurs
communication costs in partitioning the points across machines and merging
the clusters from different machines together.

\begin{figure*}
\begin{center}
\includegraphics[width=1\textwidth]{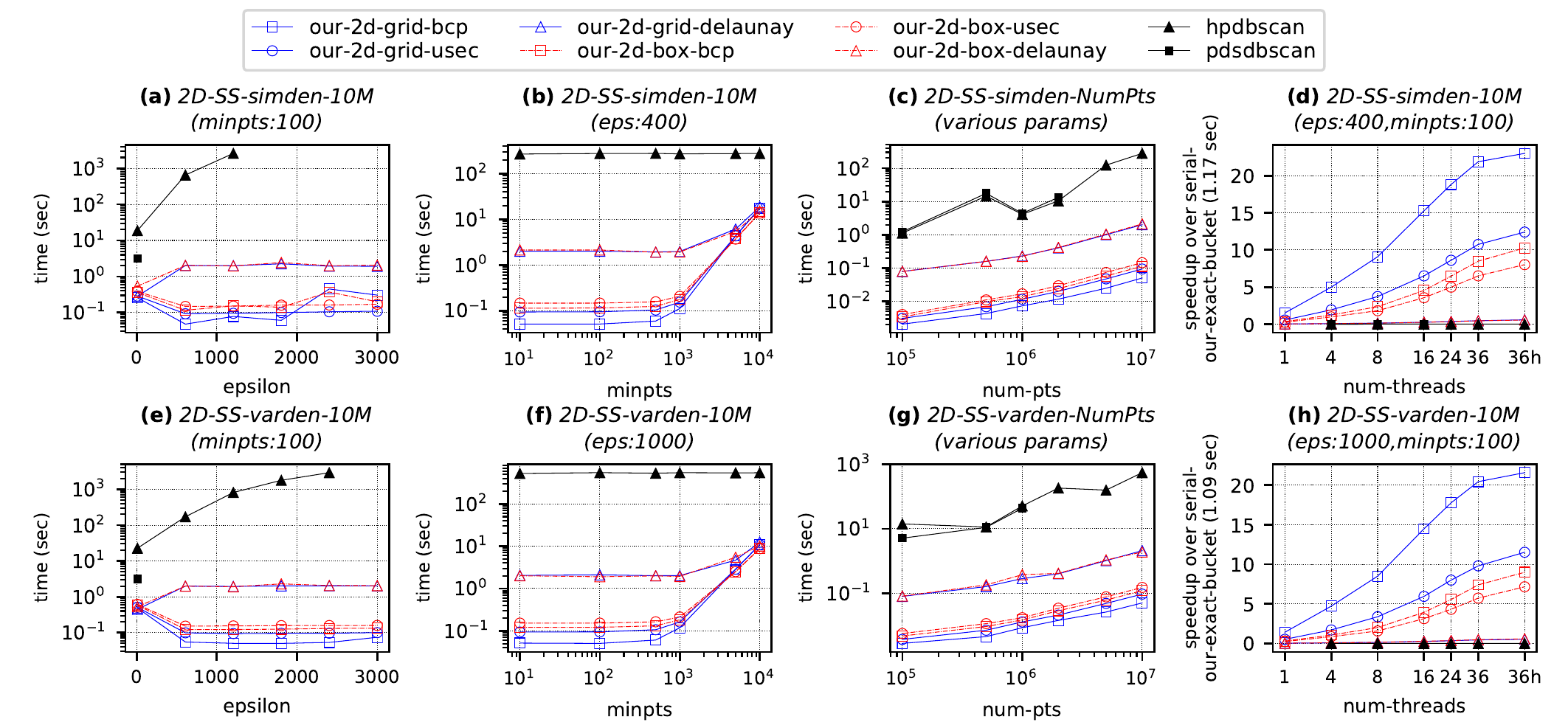}
\caption{Running time vs.  $\epsilon$, $\minpts$, number of points,
  or thread count for the 2D implementations. In (c) and (g), the
  parameters are chosen for each input size such that the algorithm
  outputs the correct clustering.  In (d) and (h), ``36h'' on the
  $x$--axis refers to 36 cores with
  hyper-threading. The $y$-axes in (a)--(c) and (e)--(g) are in log-scale.}
\label{fig:2dplots}
\end{center}
\end{figure*}

\subsection{Experiments for $d=2$}
In Figure~\ref{fig:2dplots}, we show the performance of our six 2D
algorithms as well as \emph{hpdbscan} and \emph{pdsdbscan} on the
synthetic datasets. We show the running time while varying $\epsilon$,
$\minpts$, number of points, or number of threads.  We first note that
all of our implementations are significantly faster than
\emph{hpdbscan} and \emph{pdsdbscan}.  In general, we found the
grid-based implementations to be faster than the box-based
implementations due to the higher cell construction time of the boxed-based implementations.  We also found the Delaunay
triangulation-based implementations to be significantly slower than
the BCP and USEC-based methods due to the high overhead of computing
the Delaunay triangulation. The fastest implementation overall was \emph{our-2d-grid-bcp}.


\section{Related Work}\label{sec:related}

Xu et al.\,\cite{Xu1999} provide the first parallel exact DBSCAN algorithm,
called PDBSCAN, based on a distributed $R^*$-tree.
Arlia and Coppola~\cite{Arlia2001} present
a parallel DBSCAN implementation that replicates a sequential
$R^*$-tree across machines to process points in parallel.
Coppola and Vanneschi~\cite{Coppola2002} design a parallel
algorithm using a queue to store core points, where each core point is
processed one at a time but their neighbors are checked in parallel to
see whether they should be placed at the end of the queue. Januzaj et
al.\,\cite{Januzaj2004a,Januzaj2004} design an approximate DBSCAN
algorithm based on determining representative points on different
local processors, and then running a sequential DBSCAN on the
representatives.
Brecheisen et al.\,\cite{BrecheisenKP06} parallelize a version
of DBSCAN optimized for complex distance
functions~\cite{Brecheisen2004}.

Patwary et al.\,\cite{Patwary12} present PDSDBSCAN, a multicore
and distributed algorithm for DBSCAN using a
union-find data structure for connecting points. Their union-find data
structure is lock-based whereas ours is lock-free.  Patwary et
al.\,\cite{Patwary2014,Patwary2015} also present distributed DBSCAN
algorithms that are approximate but more scalable
than PDSDBSCAN.  Hu et al.\,\cite{Hu2017} design PS-DBSCAN, an
implementation of DBSCAN using a parameter server framework.  Gotz et
al.\,\cite{Gotz2015} present HPDBSCAN, an algorithm for both
shared-memory and distributed-memory based on partitioning the data
among processors, running DBSCAN locally on each
partition, and then merging the clusters together. Very recently, Sarma et al.~\cite{Sarma2019} present a distributed algorithm, $\mu$DBSCAN, and report a running time of 41 minutes for clustering one billion 3-dimensional points using a cluster of 32 nodes. Our running times on the larger 13-dimensional TeraClickLog dataset are significantly faster (under 30 seconds on 48 cores).

Exact and approximate distributed DBSCAN
algorithms have been designed using 
MapReduce~\cite{Fu2011,Dai2012,He2014,Fu2014,Kim2014,Yu2015,Araujo2015,Hu2018} and
Spark~\cite{Cordova2015,Han2016,Luo2016,Huang2017,Song2018,Lulli2016}. RP-DBSCAN~\cite{Song2018}, an approximate DBSCAN algorithm, has been shown to be the
state-of-the-art for MapReduce and Spark.
GPU implementations of DBSCAN have also been designed~\cite{Bohm2009,Andrade2013,Welton2013,Chen2015}.


In addition to parallel solutions, there have been optimizations proposed to
speed up sequential
DBSCAN~\cite{Brecheisen2004,Kryszkiewicz2010,MaheshKumar2016}.  DBSCAN
has also been generalized to other definitions of
neighborhoods~\cite{Sander1998}.  Furthermore, there have been
variants of DBSCAN proposed in the literature, which do not return the
same result as the standard DBSCAN.  IDBSCAN~\cite{Borah2004},
FDBSCAN~\cite{Liu2006}, GF-DBSCAN~\cite{Tsai2009},
I-DBSCAN~\cite{Viswanath2006}, GNDBSCAN~\cite{Huang2009},
Rough-DBSCAN~\cite{Viswanath2009}, and DBSCAN++\,\cite{Jang2019} use
sampling to reduce the number of range queries needed.  El-Sonbaty et
al.\,\cite{El-Sonbaty2004} presents a variation that partitions the
dataset, runs DBSCAN within each partition, and merges together dense
regions. GriDBSCAN~\cite{Mahran2008} uses a similar idea with an
improved scheme for partitioning and merging. Other partitioning based
algorithms include PACA-DBSCAN~\cite{Jiang2011},
APSCAN~\cite{Chen2011}, and AA-DBSCAN~\cite{Kim2019}.  DBSCAN$^*$ and
H-DBSCAN$^*$ are variants of DBSCAN where only core points are included
in clusters~\cite{Campello2015}. Other variants use approximate
neighbor queries to speed up DBSCAN~\cite{Wu2007,HeGWW17}.

OPTICS~\cite{Ankerst1999}, SUBCLU~\cite{KroegerKK04}, and
GRIDBSCAN~\cite{Uncu2006}, are hierarchical versions of DBSCAN that
compute DBSCAN clusters on different parameters, enabling clusters of
different densities to more easily be found.
POPTICS~\cite{Patwary2013} is a parallel version of OPTICS based on
concurrent union-find.

\section{Conclusion}\label{sec:conclusion}
We have presented new parallel algorithms for exact and
approximate Euclidean DBSCAN that are both theoretically-efficient and
practical. Our algorithms are work-efficient and have polylogarithmic depth,
making them highly parallel. Our experiments demonstrate that our
solutions achieve excellent parallel speedup and significantly
outperform existing parallel DBSCAN solutions.  Future work includes
designing theoretically-efficient and practical parallel algorithms
for variants of DBSCAN and hierarchical versions of DBSCAN.

\section*{Acknowledgements} This research was supported by DOE
Early Career Award \#DE-SC0018947, NSF CAREER Award \#CCF-1845763,
Google Faculty Research Award, DARPA SDH Award \#HR0011-18-3-0007, and
Applications Driving Architectures (ADA) Research Center, a JUMP
Center co-sponsored by SRC and DARPA.

\bibliographystyle{ACM-Reference-Format}
\bibliography{references}

\appendix
\section{Proof for USEC in Section~\ref{sec:clustercore}}\label{sec:usec_proof}

\begin{figure}
  \begin{center}
    \includegraphics[width=\columnwidth]{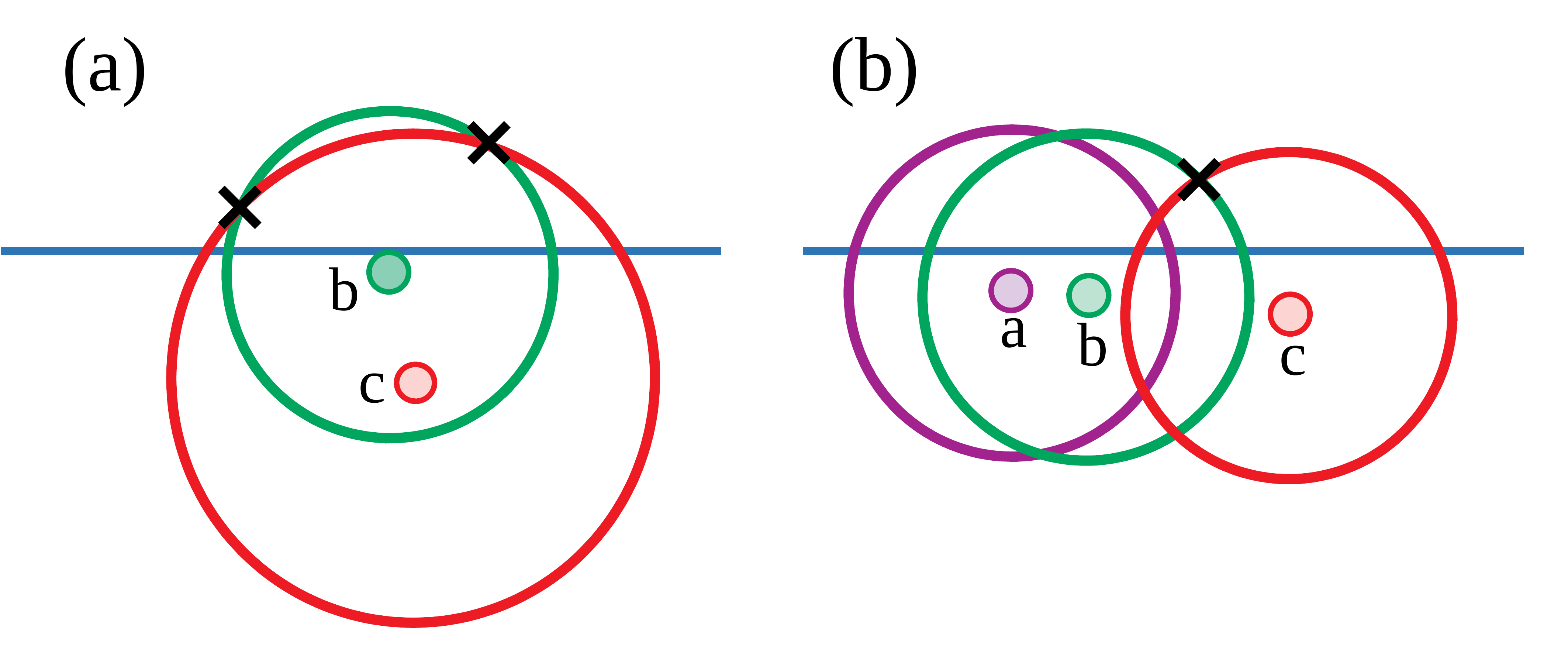}
    \caption{
      (a) shows that,
      in order for the circle centered at $c$ to intersect with the
      $\epsilon$-radius circle of $b$ at two points, the former must has a radius larger
      than $\epsilon$.
      (b) shows the circle of $c$'s intersection with that of $b$, and that
      the left half of $c$'s circle is below the union of the arcs
      formed by the circles of $a$ and $b$ above the horizontal line.
    }\label{fig:usec-proof}
  \end{center}
\end{figure}

In this section, we prove that the left and right wavefronts we merge must
intersect at one unique point. As before, we assume that the points are distinct.
Consider any three points $a$, $b$, and $c$ from the same cell, in order of $x$-coordinate.
We consider the arcs formed by their $\epsilon$-radius circles,
forming a wavefront on the top border of the cell.
We prove that the arc of $c$ can intersect at most one location in the
union of arcs formed by $a$ and $b$.
Suppose that $c$'s arc intersects at least two locations in $b$'s arc.
Take any two of these locations. They are part of the circle that forms $c$'s arc.
This is a contradiction because it implies the circle forming $c$'s arc has larger
radius than the circle forming $b$'s arc, as shown in Figure~\ref{fig:usec-proof}(a).
Therefore $c$'s arc can intersect at most one location in $b$'s arc.
A similar argument shows that $c$'s arc can intersect at most one location in $a$'s arc.

Now we need to prove that $c$'s arc cannot intersect both $a$'s arc
and $b$'s arc.  Without loss of generality, suppose that $c$'s arc
intersects with $b$'s arc.  Then, once $c$'s arc intersects with $b$'s
arc, the rest of $c$'s arc must be in the interior of $b$'s arc, and
thus below the union of $a$'s and $b$'s arcs above the horizontal line.
The rest of $c$'s arc
must be in the interior of $b$'s arc because $c$ is to the right of
$b$, and their arcs are formed by circles of the same radius. This is
shown in Figure~\ref{fig:usec-proof}(b).

Applying the above argument to all possible choices of $a$ and $b$ in
the left wavefront, and all possible choices of $c$ in the right
wavefront implies that there can only be one intersection between the
two wavefronts.

\end{document}